\documentclass[a4paper,11pt,oneside]{article}
\pdfoutput=1

\usepackage[small,bf]{caption}
\usepackage{geometry}
\usepackage{amsmath}
\usepackage{amssymb}
\usepackage{color}
\usepackage{graphicx}
\usepackage{subfig}
\usepackage[]{units}
\usepackage{booktabs}
\usepackage{verbatim}
\usepackage{float}

\newcommand{\TeV}{\,\mathrm{TeV}}
\newcommand{\GeV}{\,\mathrm{GeV}}

\newcommand{\fracwithdelims}[4]{\left#1 \frac{#3}{#4} \right#2}
\newcommand{\ord}[1]{\mathcal{O}\left( #1 \right)}

\newcommand{\VeV}[2]{#1\langle #2 #1\rangle} 

\newcommand{\Fig}[1]{Fig.~\ref{fig:#1}}

\newcommand{\eq}[1]{eq.~(\ref{eq:#1})}

\newcommand{\eqs}[1]{eqs.~(\ref{eq:#1})}

\newcommand{\nohyphens}%
        {\hyphenpenalty=10000\exhyphenpenalty=10000\relax}

\DeclareMathOperator{\tr}{Tr}

\DeclareMathOperator{\sign}{sign}

\newcommand{\GSM}{G_\text{SM}}

\allowdisplaybreaks[1]

\newlength{\myem}
\newcommand{\sep}[1]{#1}
\newcounter{mysubequation}[equation]
\renewcommand{\themysubequation}{\alph{mysubequation}}
\newcommand{\mytag}{\stepcounter{mysubequation}%
\tag{\theequation\protect\sep{\themysubequation}}}
\newcommand{\globallabel}[1]{\refstepcounter{equation}\label{#1}}

\newcommand{\SISSA}{SISSA/ISAS and INFN, I--34136 Trieste, Italy}

\newcommand{\preprintnumber}{%
SISSA  45/2012/EP}
 
\newcommand{\titletext}
{Phenomenology of Minimal Unified Tree Level Gauge Mediation at the LHC}
\newcommand{\authortext}{
\large
Maurizio Monaco$^{\, a}$,
Maurizio Pierini$^{\, b}$,
Andrea Romanino$^{\, a}$
and 
Martin Spinrath$^{\, a}$
\medskip\\\em\normalsize 
$\mbox{}^a$ \SISSA
\\[0.1\baselineskip] 
$\mbox{}^b$ CERN, CH-1211 Geneva 23, Switzerland}
\newcommand{\abstracttext}{}

\title{
\normalsize
\hspace*{\fill}
\begin{tabular}[t]{l}\preprintnumber\end{tabular}
\vspace{3\baselineskip}\\\Large\bfseries\titletext\bigskip}
\author{\begin{minipage}[t]{0.8\textwidth}
\normalsize\centering\authortext
\end{minipage}}
\date{}

\begin{document}

\bigskip
\maketitle
\begin{abstract}\normalsize\noindent
\abstracttext
We study the collider phenomenology of the minimal unified version of the supersymmetry breaking scheme called Tree-level Gauge Mediation. We identify a peculiar source of gaugino mass non-universality related to the necessary $SU(5)$-breaking in the light fermion mass ratios and a gaugino mass sum rule at the GUT scale, $3M_2 + 2M_3 = 5 M_1$, which represents a smoking gun of this scenario, together with the known tree-level sfermion mass ratio $\tilde m_{d^c,l} = \sqrt{2}\, \tilde m_{q,u^c,e^c}$. The boundary conditions of the soft SUSY breaking terms can be parameterised in terms of six relevant parameters only (plus the sign of the $\mu$-parameter). We analyze the parameter space and define three benchmark points, corresponding to the three possible NLSPs, a bino- or wino-like neutralino or the stau. The LSP is the gravitino as in gauge mediation. For these benchmark points we show possible signatures at the LHC focusing on the Razor variable. We also comment on the Higgs mass.
\end{abstract}\normalsize\vspace{\baselineskip}

\clearpage

\noindent

\tableofcontents

\newpage

\section{Introduction}

\begin{figure}
\begin{center}
\includegraphics{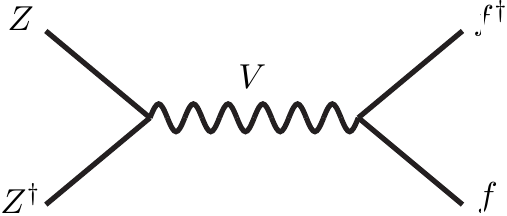}
\end{center}
\caption{The $U(1)$ massive vector superfield $V$ communicates supersymmetry breaking associated to $\VeV{\big}{Z} = F\theta^2$ to the observable field $f$ at the tree level. }
\label{fig:diagram}
\end{figure}

The aim of this paper is to spell out the LHC phenomenology of a simple, unified realization of the tree-level gauge mediation (TGM) supersymmetry breaking scheme~\cite{TGM1,TGM2}. 
In TGM, supersymmetry breaking takes place as usual in a hidden sector and is communicated at the tree level to the MSSM fields by means of superheavy vector fields associated to a broken $U(1)$ gauge group. TGM therefore requires an extension of the Standard Model (SM) gauge group $\GSM$ to at least $\GSM \times U(1)$. The (non-anomalous) extra $U(1)$ is spontaneously broken at a high scale $M$. The corresponding vector field $V$ acquires a mass $M_V = g M$, where, $g$ is the $U(1)$ gauge coupling. If both the observable superfield $f$ and the supersymmetry breaking field $Z$, $\VeV{\big}{Z} = F\theta^2$, are charged under $U(1)$, $V$ plays the role of the supersymmetry breaking messenger, as in \Fig{diagram}. At the scale $M$, the sfermion $\tilde f$ acquires a soft mass given by
\begin{equation}
\label{eq:sfermion}
\tilde m^2_f = \frac{g^2 X_f X_Z |F|^2}{M^2_V} \;,
\end{equation}
where $X_f$ and $X_Z$ are the charges of $f$ and $Z$ under $U(1)$. This simple way to communicate supersymmetry breaking is particularly suited to realize a simple, complete, and viable model of dynamical supersymmetry breaking~\cite{Caracciolo:2012de}. 

It is tempting to consider $U(1)$ as part of a grand unified group $G \supset \GSM\times U(1)$. In this case the $U(1)$-breaking scale $M$ is expected to be near the GUT scale, $M \sim M_\text{GUT}$. The minimal-rank choice for the grand unified group is then $G = \text{SO(10)}$\footnote{The other possibility, SU(6) turns out not to be phenomenologically viable.} \cite{TGM1,TGM2}, although higher rank groups such as $E_6$ can also be considered~\cite{TGME6}. The minimal choice is particularly interesting, as it gives rise to definite, peculiar predictions for the sfermion mass ratios. In this case, in fact, the $U(1)$ mediating supersymmetry breaking is associated to a well defined $SO(10)$  generator. The sfermion charges are then known up to a normalization factor, see Table~\ref{chap:TGM_LHC_tab:FieldContent}, and their squared tree-level masses, proportional to the $U(1)$ charges, are predicted (up to an overall scale factor), as in eq.~\eqref{eq:sfermion}. The embedding of the extra $U(1)$ into a Grand Unified group guarantees that the $U(1)$ quantum numbers are the same for all families (barring flavour dependent embeddings~\cite{TGM1,TGM2}), thus leading to a solution of the supersymmetric flavour problem. Note that in standard gauge mediation, a messenger scale as high as $M\sim M_\text{GUT}$ could spoil such a solution of the flavour problem, as the flavour-anarchical supergravity contributions to the soft terms would only be suppressed by a relatively mild factor $((4\pi)^2 M_\text{GUT})^2/(g^2 M_\text{Pl}))^2$, where $g^2/(4\pi)^2$ is the gauge mediation loop factor. In our case, instead, the supergravity contributions are suppressed by the much stronger factor $(M^2_\text{GUT}/M^2_\text{Pl})^2$, which is small enough to make them harmless~\cite{TGM1}. 

In this paper, we study the  LHC phenomenology of such a minimal unified setup, taking into account some notable theoretical subtleties which turn out to relate the gaugino mass ratios to the flavour structure of the SM fermions.

\section{Field content and lagrangian}

In order to study the TeV phenomenology of the model we only need to consider the lagrangian below the $SO(10)$ breaking scale. The matter field content (separated from the Higgs field content by an $R$-parity $R_P$), consists of three $\mathbf{16}_i + \mathbf{10}_i$, whose SM decomposition is given in Table~\ref{chap:TGM_LHC_tab:FieldContent}. The lower case fields are (in first approximation) the light ones. The $S_i$ are SM singlets, they may get mass at the non-renormalizable level. The other capital letter fields get mass through $SO(10)$ breaking. They consist of two pairs of vectorlike fields, $D_i^c+\overline{D_i^c}$ and $L_i+\overline{L}_i$ for each family $i=1,2,3$ and they play the role of messengers of minimal gauge mediation. We assume that only the light doublet components $h_u$, $h_d$ of the Higgs fields survive below the GUT scale (see~\cite{TGM2, doublettripletsplit} for an example of how to achieve that). If the $SO(10)$ Higgs sector contains only representations with dimension $d < 120$ ($\mathbf{10}$, $\mathbf{16} + \mathbf{\overline{16}}$, $\mathbf{45}$, $\mathbf{54}$), the doublets can only belong to $\mathbf{10}$, $\mathbf{16}$, $\overline{\mathbf{16}}$ representations. To be general, we allow them to be superpositions of the doublets in those representations. That is why their $X$ charge is not specified in Table~\ref{chap:TGM_LHC_tab:FieldContent}. The goldstino superfield $Z$ can in principle also have a (smaller) component in a $\overline{\mathbf{16}}$, see section~\ref{sec:messmass}.

\begin{table}
\centering
\begin{tabular}{ccccccc}
\toprule
Field &$SO(10)$& ${SU(3)}_C$ & ${SU(2)}_L$ & ${U(1)}_Y$ & $U(1)_X$ & $R_P$  \\ 
\midrule
$q_i$ & $\mathbf{16}_i$ & $\mathbf{3}$ & $\mathbf{2}$ & $\nicefrac{1}{6}$ & 1 & -1 \\
$u_i^c$ & $\mathbf{16}_i$ & $\mathbf{\overline{3}}$ & $\mathbf{1}$ & -$\nicefrac{2}{3}$ & 1 & -1 \\
$d_i^c$ & $\mathbf{10}_i$& $\mathbf{\overline{3}}$ & $\mathbf{1}$ & $\nicefrac{1}{3}$ & 2 & -1 \\
$l_i$ & $\mathbf{10}_i$ & $\mathbf{1}$ & $\mathbf{2}$ & -$\nicefrac{1}{2}$ & 2 & -1 \\
$e_i^c$ & $\mathbf{16}_i$ & $\mathbf{1}$ & $\mathbf{1}$ & $1$ & 1 & -1 \\
$S_i$ & $\mathbf{16}_i$ & $\mathbf{1}$ & $\mathbf{1}$ & $0$ & 5 & -1  \\
$D_i^c$ & $\mathbf{16}_i$ & $\mathbf{\overline{3}}$ & $\mathbf{1}$ & $\nicefrac{1}{3}$ & -3 & -1 \\
$\overline{D_i^c}$ & $\mathbf{10}_i$ & $\mathbf{3}$ & $\mathbf{1}$ & -$\nicefrac{1}{3}$ & -2 & -1 \\
$L_i$ & $\mathbf{16}_i$ & $\mathbf{1}$ & $\mathbf{2}$ & -$\nicefrac{1}{2}$ &  -3 & -1 \\
$\overline{L}_i$ & $\mathbf{10}_i$ & $\mathbf{1}$ & $\mathbf{2}$ & $\nicefrac{1}{2}$ & -2 & -1 \\
$h_u$ & $\mathbf{10}$, $\mathbf{\overline{16}}$, $\mathbf{\overline{16}}'$ & $\mathbf{1}$ & $\mathbf{2}$ & $\nicefrac{1}{2}$ & -- & +1 \\
$h_d$ & $\mathbf{10}$, $\mathbf{16}$, $\mathbf{16}'$ & $\mathbf{1}$ & $\mathbf{2}$ & -$\nicefrac{1}{2}$ & -- & +1 \\
$Z$ & $\mathbf{16}$ & $\mathbf{1}$ & $\mathbf{1}$ & 1 & 5 & +1 \\
\bottomrule
\end{tabular}
\caption{TGM field content. The $SO(10)$ representation to which the different superfields belong and their SM quantum numbers are shown, together with the charge under $U(1)_X$, the $SO(10)$ subgroup mediating supersymmetry breaking at the tree level, and their $R$-parity. The Higgs fields $h_u$ and $h_d$ can belong to different $SO(10)$ representations, which is why we do not specify their $U(1)_X$ charges. The field $Z$ is the source of supersymmetry breaking. \label{chap:TGM_LHC_tab:FieldContent}}
\end{table}

Whatever is the dynamics above the $SO(10)$ breaking (GUT) scale, the lagrangian below that scale can be accounted for by the most general SM and R-parity invariant lagrangian for the fields in Table~\ref{chap:TGM_LHC_tab:FieldContent}. We first give a general parameterization of the latter, which is useful to incorporate radiative corrections through RGEs, then we show how that lagrangian is determined by the few relevant parameters of the model through the boundary conditions at the GUT scale. 

The lagrangian below the GUT scale involves terms corresponding to the usual MSSM interactions and terms involving the extra heavy fields. Correspondingly, the superpotential is 
\begin{equation}
\label{eq:Wall}
W = W_{\text{MSSM}} + W_{\text{TGM}} + W_\text{S} \;,
\end{equation}
where $W_\text{S}$ depends on the singlet fields $S_i$ and is not relevant for our purposes (as long as R-parity is not spontaneously broken), and 
\begin{equation}
\label{eq:W}
\begin{aligned}
W_{\text{MSSM}} &= \lambda_U u^c q h_u + \lambda_D d^c q h_d + \lambda_E e^c l h_d + \mu h_u h_d \\
W_{\text{TGM}} &= \hat{\lambda}_D D^c q h_d + \hat{\lambda}_E e^c L h_d + M_{D} \overline{D^c} D^c + M_{D d} \overline{D^c} d^c + M_L \overline{L} L + M_{L l} \overline{L} l \,. 
\end{aligned}
\end{equation}
The terms $\overline{L}l$ and $\overline{D^c}d^c$ are supposed to be absent at the GUT scale but arise in the RGE running~\cite{doublettripletsplit}, as shown in the appendix \ref{app:TGM_muparameter}. The SUSY breaking lagrangian is
\begin{equation}
\label{eq:LSB}
\mathcal{L}_{\text{SB}} = \mathcal{L}^A_{\text{MSSM}} + \mathcal{L}^A_{\text{TGM}} + \mathcal{L}^m_{\text{MSSM}} + \mathcal{L}^m_{\text{TGM}} + \mathcal{L}^{g}_{\text{MSSM}} \;,
\end{equation}
with 
\begin{equation}
\begin{aligned}
\label{eq:LSBdetail}
- \mathcal{L}^A_{\text{MSSM}} &= A_U \tilde{u}^c \tilde{q} h_u + A_D \tilde{d}^c \tilde{q} h_d + A_E \tilde{e}^c \tilde{l} h_d + B h_u h_d + \text{h.c.} \\[1mm]
- \mathcal{L}^A_{\text{TGM}} &= \hat{A}_D \tilde{D}^c \tilde{q} h_d + \hat{A}_E \tilde{e}^c \tilde{L} h_d + B_{D} \tilde{\overline{D^c}} \tilde{D}^c + B_{D d} \tilde{\overline{D^c}} \tilde{d}^c + B_L \tilde{\overline{L}} \tilde{L} + B_{L l} \tilde{\overline{L}} \tilde{l} + \text{h.c.}  \\[1mm]
- \mathcal{L}^m_{\text{MSSM}} &= m^2_{h_u} h_u^{\dagger} h_u + m^2_{h_d} h_d^{\dagger} h_d + m^2_{q} \tilde{q}^{\dagger} \tilde{q} +  m^2_{u^c}  {\tilde{u}^{c \dagger}} \tilde{u}^c 
 + m^2_{l} \tilde{l}^{\dagger} \tilde{l}  + m^2_{d^c} {\tilde{d}^{c \dagger}} \tilde{d}^c + m^2_{e^c} {\tilde{e}^{c \dagger}} \tilde{e}^c 
 \\
- \mathcal{L}^m_{\text{TGM}} &= m^2_{D^c} {\tilde{D}^{c \dagger}} \tilde{D}^c + m^2_{\overline{D^c}} {\tilde{\overline{D^c}}}^{\dagger} \tilde{\overline{D^c}} + m^2_{L} {\tilde{L}}^{\dagger} \tilde{L} + m^2_{\overline{L}} {\tilde{\overline{L}}}^{\dagger} \tilde{\overline{L}} + (m^2_{D d} {\tilde{D}^{c \dagger}} \tilde{d}^c + m^2_{L l} {\tilde{L}}^{\dagger} \tilde{l} + \text{h.c.}) \\ 
- \mathcal{L}^{g}_{\text{MSSM}} &= \frac{1}{2} M_a \lambda_a \lambda_a + \text{h.c.} \,.
\end{aligned}
\end{equation}
In the above equations we have suppressed the flavour indexes. The terms including the supersymmetry breaking source $Z$ have also been omitted, but we will discuss them in  section~\ref{sec:messmass}.

\section{The parameters of the model}

In this section, we define the parameters of the model and show how they determine the lagrangian at the GUT scale. The TeV-scale lagrangian will then be obtained as usual by RGE running, for which we provide analytical formulas and a numerical implementation in {\tt softSUSY}~\cite{Allanach:2001kg}. 

The section is divided in two parts. In section~\ref{sec:RelevantParameters} we collect and discuss the relevant parameters of the model. This first part contains all the information needed for the  phenomenological analyses in the subsequent sections~\ref{sec:analysis} and~\ref{sec:razor}. In the remainder of the section, we discuss the details of the determination of the spectrum in terms of those parameters (and others), in particular the generation of sizeable $A$-terms due to the built-in matter-messenger couplings. This second part can be skipped on first reading. 

\subsection{Relevant parameters}
\label{sec:RelevantParameters}

Let us discuss the parameters that essentially determine the spectrum. They are: 
\begin{equation}
\label{eq:parameters}
m_{10} \;,\quad
\theta_u \; ,\quad \theta_d \;,\quad M_{1/2} \;,\quad r \;,\quad
\tan\beta \;,\quad \sign(\mu) \;.
\end{equation}
Additional parameters are involved in the determination of the detailed flavour structure of the lagrangian in eqs.~(\ref{eq:W}--\ref{eq:LSBdetail}), but they have a marginal effect on the TeV spectrum. We will discuss them in Section~\ref{sec:messmass}. 

Let us discuss the parameters in eq.~\eqref{eq:parameters} in turn. The parameter $m_{10}$ is the common tree-level mass of the MSSM sfermions belonging to a \textbf{10} of $SU(5)$, $\tilde q$, $\tilde u^c$, and $\tilde e^c$. All sfermion masses are determined (at the tree level) by $m_{10}$ through eq.~\eqref{eq:sfermion}:
\begin{equation}
\label{eq:sfermionmasses}
\tilde m^2_q = \tilde m^2_{u^c} = \tilde m^2_{e^c} = m^2_{10} \;,\quad
\tilde m^2_l = \tilde m^2_{d^c} = 2 \, m^2_{10} \;, \quad\text{with }m^2_{10} = \frac{1}{10}\frac{F^2}{M^2}\;.
\end{equation}
The factor $2$ is a prediction of the minimal unified realization of TGM. It arises because the squared sfermion masses are proportional to their charges under the $U(1)$ mediating supersymmetry breaking (see Table~\ref{chap:TGM_LHC_tab:FieldContent}). The sfermion masses originate at the scale $M$, which we identify with $M_\text{GUT}$. Here and in the following we will assume that $F/M$ is real.

The $D^c$, $\overline{D^c}$, $L$, and $\overline{L}$ soft masses are subdominant with respect to the much larger supersymmetric masses $M_D$, $M_L$ in the superpotential and, as the parameters $m^2_{Dd}$, $m^2_{Ll}$, are not relevant in our results. For completeness, they are given at the GUT scale by 
\begin{equation}
\label{eq:smessengermasses}
m^2_{D_c} = m^2_{L} = -3 m^2_{10}, \quad
m^2_{\overline{D_c}} = m^2_{\overline{L}} = -2 m^2_{10} \,, \quad
m^2_{Dd} = m^2_{Ll} = 0 .
\end{equation}

The angles $0 \leq \theta_u, \theta_d \leq \pi/2$ account for the possibility that the light MSSM Higgs $h_u$ and $h_d$ are superpositions of doublets in different $SO(10)$ representations. Given the embedding of MSSM fields in Table~\ref{chap:TGM_LHC_tab:FieldContent} (and up to non-renormalizable contributions), the up and down quark Yukawa couplings $\lambda_U u^c q h_u$ and $\lambda_D d^c q h_d$ in eq.~\eqref{eq:W} must come from the $SO(10)$ interactions $\mathbf{16}\, \mathbf{16}\, \mathbf{10}_H$ and $\mathbf{10}\, \mathbf{16}\, \mathbf{16}_H$ respectively\footnote{We can assume without loss of generality that $\mathbf{10}_H$ is the only $\mathbf{10}$ representation of $SO(10)$ containing $h_u$ and $\mathbf{16}_H$ is the only $\mathbf{16}$ representation of $SO(10)$ containing $h_d$.}. Therefore, $h_u$ must have a component in $\mathbf{10}_H$ and $h_d$ must have a component in $\mathbf{16}_H$. The simplest possibility is that this is it. On the other hand, to be general, we can consider the possibility that $h_u$ has also a component in a $\overline{\mathbf{16}}$ and $h_d$ in a $\mathbf{10}$ (there are no further possibilities as we only consider $SO(10)$ representations with dimension $d < 120$). In this case we use the angles $\theta_u$ and $\theta_d$ to measure the size of the Higgs components in the different representations:
\begin{equation}
\label{eq:higgscomponents}
\mathbf{10}_H \supset \cos\theta_u h_u + \ldots \qquad \mathbf{16}_H \supset \sin\theta_d h_d + \ldots .
\end{equation}
In the ``pure'' case in which the light Higgs doublets are contained in the $\mathbf{10}_H$ and $\mathbf{16}_H$ only, their $U(1)_X$ charges are given by: $X_{h_u} = -( X_q + X_{u^c}) = -2$, $X_{h_d} = -( X_q + X_{d^c}) = -3$. The charges are negative because the MSSM Yukawas must be $U(1)_X$ invariant and the sfermions must have positive charges. Their soft masses are therefore negative at the tree level. In the general case, we have instead
\begin{equation}
m_{h_u}^2 = (-2 \cos^2 \theta_u + 3 \sin^2 \theta_u)\, m_{10}^2 \quad \text{and} \quad m_{h_d}^2 = (2 \cos^2 \theta_d - 3 \sin^2 \theta_d)\, m_{10}^2 \label{eq:HiggsBC}
\end{equation}
and the soft masses can both be positive or negative at the tree level. 

\begin{figure}
 \centering
\includegraphics[width=0.50\textwidth]{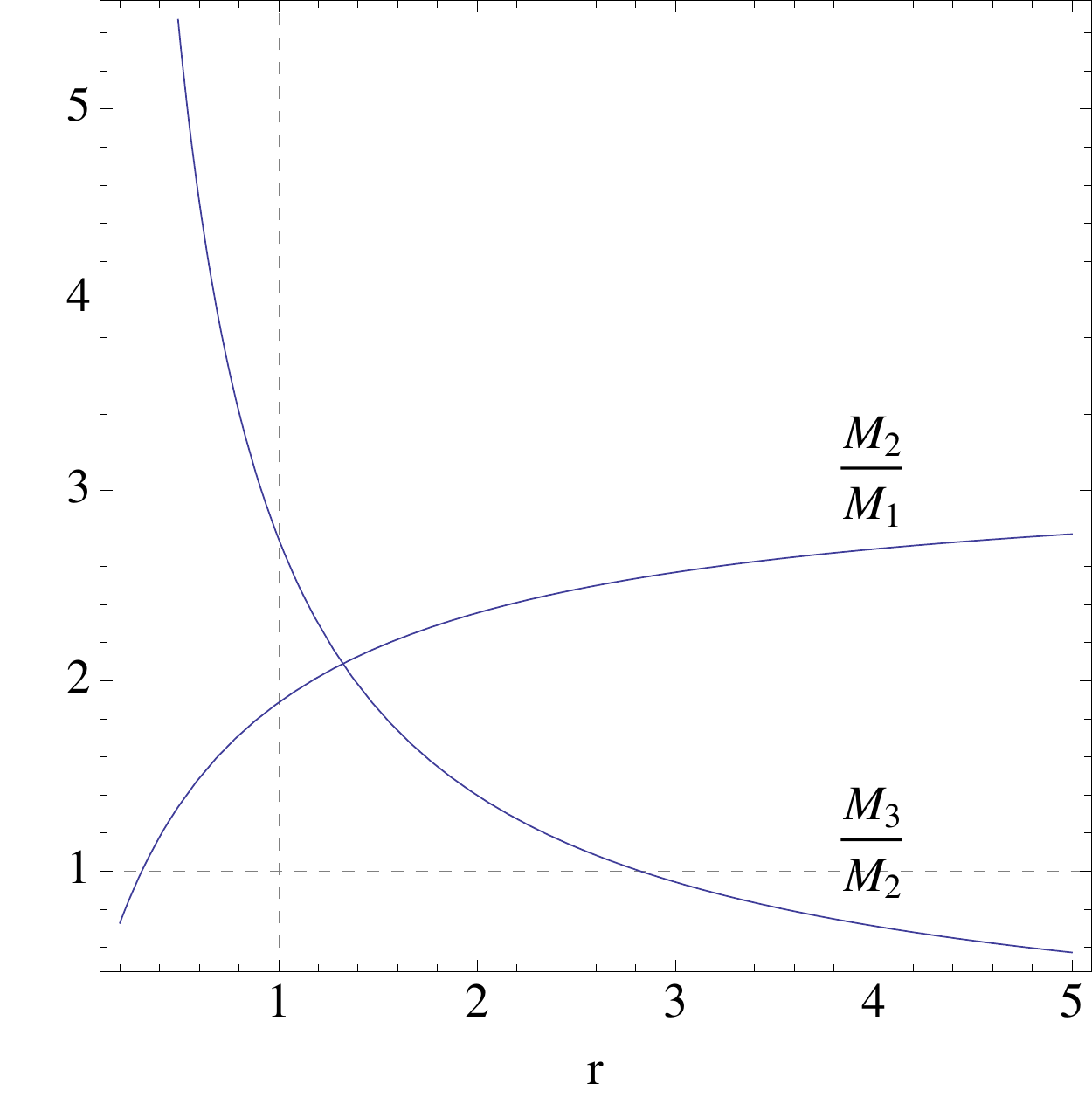}
 \caption{Dependence of the gaugino mass parameter ratios $M_2/M_1$ and $M_3/M_2$ evaluated at the SUSY breaking scale on the parameter $r$. The wino mass term $M_2$ is lighter than the bino mass term $M_1$ for $r\lesssim 0.3$.}
\label{fig:gauginos}
\end{figure}

The gaugino masses are generated at the one-loop level by the couplings of $D^c$, $\overline{D^c}$, $L$, $\overline{L}$, which act as messengers of minimal gauge mediation, to the supersymmetry breaking source~\cite{TGM1,TGM2}. They are determined in terms of the parameters $M_{1/2}$ and $r$ according to
\begin{equation}
\label{eq:trade}
M_{1/2} = \frac{M_2+M_3}{2}, \quad r = \frac{M_2}{M_3} \quad \text{(GUT scale)} \;,
\end{equation}
with $M_1$ given by the sum rule 
\begin{equation}
\label{eq:sumrule}
M_1 = \frac{3}{5} M_2 + \frac{2}{5} M_3  \quad \text{(GUT scale)} \;.
\end{equation}
Note that $r=1$ corresponds to universal gaugino masses. Largely non universal masses can arise for $r\neq 1$, despite $SO(10)$ unification, as will be discussed in more detail in section \ref{sec:gauginos}. As a consequence, i) small values of $r$ can make the Wino lighter than the Bino and ii) the measurement of non-universal gaugino masses satisfying the sum rule~(\ref{eq:sumrule}) can be considered as another smoking gun of minimal unified TGM. The dependence of the gaugino mass parameter ratios $M_2/M_1$ and $M_3/M_2$ on $r$ at the SUSY breaking scale scale is shown in figure \ref{fig:gauginos}.

As usual, $\tan\beta$ can be traded for the $B$ parameter in eq.~\eqref{eq:LSBdetail} and $\text{sign}(\mu)$, together with the EWSB condition, determine the $\mu$ parameter.

Eqs.~(\ref{eq:sfermionmasses},\ref{eq:smessengermasses},\ref{eq:HiggsBC}--\ref{eq:sumrule}) determine the boundary conditions of all the soft parameters except the MSSM $A$-terms and the parameters of $\mathcal{L}^A_\text{TGM}$. The latter, as the heavy field soft terms, are not relevant in our results. The MSSM $A$-terms are instead of course relevant. Usually in gauge mediation it is assumed that the visible sector has only gauge interactions with the hidden sector and hence no $A$-terms are generated at the one-loop level. This is not the case in unified TGM, where the MSSM fields and the minimal gauge mediation messengers lie in the same $SO(10)$ multiplets, so that the messenger-messenger-$Z$ coupling generating gaugino masses are accompanied by matter-messenger-$Z$ couplings generating non-vanishing $A$-terms at the messenger scale. Such $A$-terms are rather model-dependent, as their values depend in the detailed form of the $SO(10)$ lagrangian and on the implementation of doublet-triplet splitting in the Higgs sector. We will specify the prescription we use for the $A$-terms in section~\ref{sec:Aterms}.

\subsection{Heavy chiral messengers and marginal parameters}
\label{sec:messmass}

In this subsection and in the next ones, we provide the details of the determination of the spectrum in terms of the relevant parameters and introduce additional physical parameters that have a marginal effect on the spectrum. 

Let us begin with the scale $M$ at which the $U(1)_X$ subgroup of $SO(10)$ is broken and the sfermion masses are generated, and their RGE evolution begins, which is expected to lie near the GUT scale. The TeV-scale predictions have only a mild (logarithmic) dependence on the precise value of $M$. We therefore set $M = M_\text{GUT}$ in our numerical results. 

The spectrum below the GUT scale contains the MSSM fields and the extra heavy fields $D^c+\overline{D^c}$ and $L+\overline{L}$. Such fields play an important role in the determination of the TeV-scale lagrangian. In fact, their coupling to supersymmetry breaking generates gaugino masses at the one loop. Moreover, their presence at high scale affects the running of the MSSM parameters. In order to compute the low energy spectrum, it is therefore necessary to know their masses and their couplings to supersymmetry breaking and MSSM fields. 

Since the $D^c,L$ and $\overline{D^c},\overline{L}$ fields belong to different $SO(10)$ representations, they acquire masses through $SO(10)$ breaking, specifically through the vev of the SM singlet components of a $\mathbf{16}+\overline{\mathbf{16}}$, denoted by $M > 0$\footnote{The $D$-term condition for the $U(1)_X$ forces the two vevs to be equal in absolute value, up to negligible SUSY breaking effects. $M$ can be taken positive without loss of generality.}. We expect $M$ to be of the order of the GUT scale, $M\sim M_\text{GUT}$ and denote by $r_\text{GUT} \equiv M/M_\text{GUT}$ their $\ord{1}$ ratio. It is therefore convenient to write the mass terms in eq.~\eqref{eq:W} as 
\begin{equation}
\label{eq:WM}
M^D_{ij} \overline{D^c}_i D^c_j + M^L_{ij} \overline{L}_i L_j = h^D_{ij} M \overline{D^c}_i D^c_j +  h^L_{ij} M \overline{L}_i L_j \,.
\end{equation}
The couplings $h_D$, $h_L$ arise from the $SO(10)$ superpotential~\cite{TGM1,TGM2}
\begin{equation}
\label{eq:W2}
W_2 = h_{ij} \mathbf{16}_i \mathbf{10}_j \mathbf{16} + h'_{ij} \mathbf{16}_i \mathbf{10}_j \mathbf{16}' + \frac{y_{ij}}{2} \mathbf{16}_i \mathbf{16}_j \mathbf{10} + W_2^\text{NR}
\end{equation}
after substituting the vevs of the $\mathbf{16}$. In the $SO(10)$ limit, $h_D = h_L = h$. Corrections to the above relations can originate from the non-renormalizable part of the superpotential, $W_2^\text{NR}$. From \eq{W2} it also follows that the mixing parameters $M_{dD}$ and $M_{lL}$ in eq.~\eqref{eq:W} vanish at the GUT scale at the renormalizable level, and are therefore set to zero at that scale:
\begin{equation}
\label{eq:mixingpars}
M_{dD} = 0, \qquad  M_{lL} = 0 \qquad \text{(GUT scale)} \,.
\end{equation}
Non vanishing values are generated by the RGE running between the GUT and the messenger scales, as no unbroken quantum number distinguishes the $d^c,l$ fields from the $D^c,L$ ones. 

Supersymmetry breaking must be provided by the $F$-term vev of the SM singlet component of spinorial representations of $SO(10)$, which are however forced by gauge invariance not to coincide with $\mathbf{16}$, $\overline{\mathbf{16}}$ (see the discussion in \cite{TGM2}) and will therefore be denoted by $\mathbf{16}^\prime$, $\overline{\mathbf{16}}^\prime$. In order to obtain positive tree level sfermion masses, the $F$-term of the $\mathbf{16}^\prime$ must be larger than the one of the $\overline{\mathbf{16}}^\prime$ \cite{TGM2}. We will then assume for simplicity that only the SM singlet component of the $\mathbf{16}^\prime$ field, $Z$, gets an $F$-term vev $F$. As $|F| \ll M^2$, the field $Z$ should be included in the effective lagrangian below the GUT scale defined by equations~(\ref{eq:Wall}--\ref{eq:LSBdetail}). The relevant terms are the superpotential couplings
\begin{equation}
\label{eq:WZ}
W_Z = h^{\prime D}_{ij} Z \overline{D^c}_i D^c_j +  h^{\prime L}_{ij} Z \overline{L}_i L_j \,.
\end{equation}
The couplings $h'_D$, $h'_L$ arise from the $SO(10)$ superpotential in \eq{W2} after substituting the $F$-term vev of $\mathbf{16}'$. In the $SO(10)$ limit, $h'_D = h'_L = h'$. For simplicity, we will neglect the flavour structure of the matrices $h_D$, $h_L$, $h'_D$, $h'_L$ and consider only the diagonal elements, assuming that, as in the case of the SM Yukawa couplings, the deviation from the diagonal form, i.e.\ the breaking of the individual flavour numbers, is small. In such a case, the flavour structure we are neglecting does not significantly affect the collider observables we are interested in. Eqs.~\eqref{eq:WM} and \eqref{eq:WZ} then involve six new parameters each. The latter are related to the MSSM down quark and charged lepton Yukawa couplings, as discussed in detail in appendix \ref{sec:parameters}. As shown there, the simplest possible prediction for the messenger mass parameters $h_{D,L}$, which arises in the hypothesis of minimal Higgs embedding, $\mathbf{16}_H = \mathbf{16}$, is that they are proportional to the corresponding SM Yukawa couplings (in the presence of heavy Higgs triplets, this hypothesis gives rise to a predictive scheme for leptogenesis~\cite{Frigerio:2008ai}): 
\begin{equation}
\label{eq:minimalpars}
h_D = \lambda_D / \sin\theta_d \qquad
h_L = \lambda_E / \sin\theta_d .
\end{equation}
We expect in this case the couplings $h_{D_i,L_i}$, and therefore the messenger masses, to follow the same hierarchy as the corresponding fermion masses, with the first two family of messenger significantly lighter than the third one. The prediction in \eq{minimalpars}, however, can receive corrections if the light Higgs fields have also a component in the $\mathbf{16}'$, $\overline{\mathbf{16}}'$. Moreover, the $SO(10)$ relations between SM fermion and messenger couplings in \eqs{minimalpars} might receive corrections from the same sources of $SO(10)$ breaking needed to fix the  GUT prediction for the light fermion mass ratios, i.e.\ to differentiate $\lambda_D$ and $\lambda^T_E$. In order to be general, we therefore modify the relations in \eqs{minimalpars} by introducing new parameters $c_{D_i}$, $c_{L_i}$, $i=1,2,3$,
\begin{equation}
\label{eq:cpars}
h^D_i = c_{D_i} \lambda^D_i / \sin\theta_d 
\qquad
h^L_i = c_{L_i} \lambda^L_i / \sin\theta_d ,
\end{equation}
whose relation with the fundamental parameters of the theory is discussed in Appendix~\ref{sec:parameters}. The choice $c_{D_i,L_i} = 1$ in \eq{cpars} corresponds to the minimal setting in \eqs{minimalpars}. We have checked that our TeV scale predictions have a very mild (logarithmic) dependence on $\ord{1}$ variations of the parameters $c_{D_i,L_i}$. Therefore, we set them to a reference value of 0.1 in most of the numerical results below, while keeping the possibility to give them an arbitrary value in our codes. 

As for the couplings to supersymmetry breaking $h'_{D_i}$ and $h'_{L_i}$, they are conveniently traded for the parameters $\gamma_{D_i}$, $\gamma_{L_i}$ defined by
\begin{equation}
\label{eq:gammas}
\gamma_{D_i} \equiv \fracwithdelims{(}{)}{h^{\prime}_{D_i}}{h_{D_i}}_{M_{D_i}}
\qquad
\gamma_{L_i} \equiv \fracwithdelims{(}{)}{h^{\prime}_{L_i}}{h_{L_i}}_{M_{L_i}} ,
\end{equation}
where the couplings are supposed to be evaluated at the corresponding heavy field mass scale $M_{D_i} = h_{D_i} M$, $M_{L_i} = h_{L_i} M$. In the next section we will show how the above parameters enter the determination of gaugino masses. For the time being, it suffices to note that the 6 parameters $\gamma_{D_i}$, $\gamma_{L_i}$, and therefore the couplings $h'_{D_i}$, $h'_{L_i}$, are determined in terms of $M_{1/2}$, $r$ (which determine, as we will see, the two averages $\gamma_D \equiv (\sum_{i=1}^3 \gamma_{D_i})/3$ and  $\gamma_L \equiv (\sum_{i=1}^3 \gamma_{L_i})/3$), and the four ratios
\begin{equation}
\label{eq:rLD}
r_{D_i} = \gamma_{D_i}/\gamma_{D}, \quad  r_{L_i} = \gamma_{L_i}/\gamma_{L}, \quad i=1,2 . 
\end{equation}
Again, the four parameters $r_{D_i},r_{L_i}$ can be expected to be of order one and we have checked that our TeV scale predictions have a very mild dependence on $\ord{1}$ variations of those parameters. Therefore, we set them to 1, unless otherwise stated, in the numerical results below, while keeping the possibility to give them an arbitrary value in our codes. 

The Yukawa couplings $\hat{\lambda}_D$ and $\hat{\lambda}_E$ in \eq{W} are related to the up-type quark Yukawa couplings by the relations
\begin{equation}
 \hat{\lambda}_D = \hat{\lambda}_E = \frac{\cos \theta_d}{\cos \theta_u} \lambda_U 
\end{equation}
at the GUT scale, where we neglected possible contributions from non-renormalizable operators, as they are liklely to only affect the small couplings of the first two families, which are not relevant for our purposes. 

To sum up, in this section we have specified the GUT scale boundary conditions for all the parameters in  \eq{W}. The Yukawas $\lambda_U$, $\lambda_D$, $\lambda_E$ are determined at low energy by the SM fermion masses, and $\hat{\lambda}_D$, $\hat{\lambda}_E$ from GUT scale relations. The messenger masses $M_{D,L}$ are specified by eqs.~(\ref{eq:WM},\ref{eq:cpars}) while the parameters $M_{dD}$, $M_{lL}$ are set to zero at the GUT scale. The $\mu$ parameter is determined by the EWSB condition and the specification of its sign. The parameters in \eq{WZ} will be determined in the next section by eqs.~(\ref{eq:gammas},\ref{eq:rLD},\ref{eq:gamma},\ref{eq:M12}) together with \eq{trade}.

\subsection{Gaugino masses in greater detail}
\label{sec:gauginos}

In the remainder of this section, we will discuss in greater detail the determination of the soft terms. 

As mentioned, gaugino masses are generated, as in minimal gauge mediation (MGM), at the one loop level because of the coupling of the supersymmetry breaking field $Z$ to the heavy $D^c$, $\overline{D^c}$, $L$ and $\overline{L}$ fields, which play the role of chiral messengers of SUSY breaking. 

While in MGM both the supersymmetric and supersymmetry breaking messenger masses come from the same Yukawa couplings, the ones to the spurion, here they are associated to two independent sets of couplings, the ones to $U(1)_X$ breaking, $h_{D,L}$, and the ones to supersymmetry breaking, $h'_{D,L}$. This opens the possibility to enhance gaugino masses by means of the ratio of the couplings. In this section we show how such features are implemented in the $SO(10)$ model under consideration, taking into account possible $SO(10)$ breaking effects, and we point out a possible source of non minimality of gaugino masses, accounted for by the parameter $r$ in \eq{parameters}. 

Gaugino masses can be expressed in terms of the messenger masses and couplings to SUSY breaking in eqs.~\eqref{eq:WM} and \eqref{eq:WZ}. The six vectorlike chiral messengers, $\overline{D_i^c} + D^c_i$ and $\overline{L}_i + L_i$, $i=1,2,3$, have masses $M_{D_i} = {h_D}_i M$ and $M_{L_i} = {h_L}_i M$ respectively. Their scalar components get supersymmetry breaking mass terms given by ${h_D^\prime}_i F$ and ${h_L^\prime}_i F$. The contributions of the $i$-th family of messengers to the gaugino masses $M_a$, $a = 1,2,3$, are then
\begin{equation}
\label{eq:gauginos}
M^{D_i}_a = \frac{\alpha_a(M_{D_i})}{4\pi} b^D_a {\gamma_D}_i \frac{F}{M}
\quad\text{(scale $M_{D_i}$)} \quad
M^{L_i}_a = \frac{\alpha_a(M_{L_i})}{4\pi} b^L_a {\gamma_L}_i \frac{F}{M}
\quad\text{(scale $M_{L_i}$)} ,
\end{equation}
where $b^D = (2/5,0,1)$, $b^L = (3/5,1,0)$,  and the parameters ${\gamma_D}_i,{\gamma_L}_i$ are defined in eq.~\eqref{eq:gammas}. Each of those contributions arise at the scale of the corresponding messenger and the gauge couplings in eqs.~\eqref{eq:gauginos} are supposed to be evaluated at that scale, which is different for each contribution. The individual contributions in eqs.~\eqref{eq:gauginos} can be formally obtained from the one loop running from the GUT scale of the hypothetical values
\begin{equation}
\label{eq:gauginos2}
M^{D_i}_a = \frac{\alpha_a(M_\text{GUT})}{4\pi} b^D_a {\gamma_D}_i \frac{F}{M}
\qquad
M^{L_i}_a = \frac{\alpha_a(M_\text{GUT})}{4\pi} b^L_a {\gamma_L}_i \frac{F}{M}
\quad\text{(GUT scale)} ,
\end{equation}
where now the gauge couplings are supposed to be evaluated at the GUT scale, while $\gamma_D$ and $\gamma_L$ are still given by eq.~\eqref{eq:gammas}. At the GUT scale, the individual contributions in eqs.~\eqref{eq:gauginos2} can be summed to give
\begin{equation}
\label{eq:gauginos3}
M_a = 3 \frac{\alpha_a(M_\text{GUT})}{4\pi} \left(
2\frac{b^D_a + r b^L_a}{1+r} 
\right)\gamma \frac{F}{M}
\quad\text{(GUT scale)} ,
\end{equation}
where $r$ is a ratio and $\gamma$ is the average of the six parameters defined in eq.~\eqref{eq:gammas}:
\begin{equation}
\label{eq:gamma}
r = \frac{\sum_{i=1}^3 {\gamma_L}_i}{\sum_{i=1}^3 {\gamma_{D}}_i}\,, \qquad
\gamma = \frac{1}{6} \bigg(
\sum_{i=1}^3 {\gamma_D}_i + \sum_{i=1}^3 {\gamma_{L}}_i
\bigg)\, .
\end{equation}
We can conveniently trade the parameter $\gamma$ in terms of the more useful\footnote{If gauge couplings do not unify one should use $(\alpha_2(M_\text{GUT}) + \alpha_3(M_\text{GUT}))/2$ instead of $\alpha_\text{GUT}$.}
\begin{equation}
\label{eq:M12}
M_{1/2} \equiv 3\, \frac{\alpha_\text{GUT}}{4\pi}\gamma \frac{F}{M} \,,
\end{equation} 
and thus obtain the parameterization of gaugino masses in terms of $M_{1/2}$ and $r$ in  \eqs{trade} and the sum rule in \eq{sumrule}. As stressed above, those relations are valid at the GUT scale only in the sense that the gaugino masses at the scales at which they are actually generated and below can be obtained by running the formal GUT scale values with one loop RGEs. 

The gaugino mass parameters $M_{1/2}$ can well be of the order of the tree level stop mass $m_{10}$, despite it is generated at the one loop level \cite{TGM1}. This is in part due to the fact that $F/M = \sqrt{10}\, m_{10}$, giving a factor $3 \sqrt{10}$ enhancement of the loop suppressed value
\begin{equation}
\label{eq:M12m10}
M_{1/2} \equiv  \frac{\alpha_{\text{GUT}}}{4\pi} (3 \sqrt{10} \,\gamma) m_{10} \,.
\end{equation}
And it is in part due to the fact that the unknown factor $\gamma$, being essentially a ratio of presumably hierarchical Yukawa couplings, can easily be larger (or smaller) than 1. 

The gaugino masses obtained in this way are potentially non universal at the GUT scale, without any conflict with gauge coupling unification, if the parameter $r$ is different from 1. Let us close this section by discussing how concrete is such a possibility. The $SU(5)$ gauge symmetry, if unbroken, would force ${\gamma_D}_i = {\gamma_L}_i$ and $r=1$. On the other hand, the possibility that $r\neq 1$ is plausible because $SU(5)$ is broken and the same $SU(5)$ breaking corrections needed to make $\lambda_D \neq \lambda_E$ can as well make $ h_D \neq h_L$ and $h'_D \neq h'_L$, so that ${\gamma_D}_i \neq {\gamma_L}_i$ and $r \neq 1$. Note that even in the limit in which the $SU(5)$ breaking effects are small and only affect significantly the small Yukawa couplings of the first families, the effect on $r$ can be sizeable. In fact, the ratio of the small Yukawa couplings, potentially significantly different from 1, enters the $r$ parameter with the same weight as the ratio of the third family Yukawas. 

\subsection{Trilinear terms}
\label{sec:Aterms}

The MSSM trilinear terms in eq.~\eqref{eq:LSBdetail} are generated through one loop graphs at the scale at which the heavy $D^c$, $\overline{D^c}$, $L$ and $\overline{L}$ are integrated out. In the region of the parameter space where the messenger masses are well below the GUT scale, the loops generating the $A$-terms are dominated by the contribution of the messengers, with the contribution of fields living at the GUT scale suppressed by their higher mass. In such a scase, the trilinears have the following form 
\begin{equation}
\label{eq:trilinear101}
\begin{aligned}
A_U &= A_{u^c} \lambda_U + \lambda_U A_q + \lambda_U A_{h_u} \;, \\
A_D &= A_{d^c} \lambda_D + \lambda_D A_q + \lambda_D A_{h_d}  \;,\\
A_E &= A_{e^c} \lambda_E + \lambda_E A_l  + \lambda_E A_{h_d} \;.
\end{aligned}
\end{equation}
More precisely, the contributions induced by the coloured messengers $D^c$ and $\overline{D^c}$ are 
\globallabel{eq:trilinear102}
\begin{align}
A_q ({M_D}_i) &=  - \frac{1}{(4 \pi)^2} {\gamma_D}_i   {\hat{\lambda}_{D_i}}^2 \,   \frac{F}{M}  \;,\mytag \\
A_{h_d} ({M_D}_i) &= - \frac{3}{(4 \pi)^2} {\gamma_D}_i  {\hat{\lambda}_{D_i}}^2\,    \frac{F}{M}   \;, \mytag\\[1mm]
A_l ({M_D}_i) &= A_{d^c} ({M_D}_i) = A_{u^c} ({M_D}_i) = A_{e^c} ({M_D}_i) = A_{h_u} ({M_D}_i) = 0 \;,\mytag
\end{align}
while the one induced by $L$ and $\overline{L}$ are
\globallabel{eq:trilinear103}
\begin{align}
A_{e^c} ({M_L}_i) &=  - \frac{2}{(4 \pi)^2} {\gamma_L}_i   {\hat{\lambda}_{E_i}}^2 \,  \frac{F}{M}  \;,\mytag \\
A_{h_d} ({M_L}_i) &= - \frac{1}{(4 \pi)^2} {\gamma_L}_i  {\hat{\lambda}_{E_i}}^2\,    \frac{F}{M}   \;, \mytag\\[1mm]
A_l ({M_L}_i) &= A_{d^c} ({M_L}_i) = A_{u^c} ({M_L}_i) = A_{q} ({M_L}_i) = A_{h_u} ({M_L}_i) = 0 \;. \mytag
\end{align}
Note that only the third family $A$-terms are non negligible, as the first and second family ones are suppressed by powers of small Yukawa couplings. This solves the supersymmetric CP problem. 

On the other hand, if $\tan\beta$ is largish and/or $\theta_d$ is small, the third family messenger masses can be close to the GUT scale. This possibility is particularly interesting, as it corresponds to third family Yukawa couplings of order 1 in the microscopic theory at the GUT scale. In fact, let us remind that $M_{D_i,L_i} \sim h_{D_i,L_i} M \sim h_{D_i,L_i} M_\text{GUT}$, with the third family expected to be largest. Therefore, $M_{D_3,L_3} \sim M_\text{GUT}$ requires $h_{D_3,L_3}\sim 1$. In such a case, the suppression of the bottom and the tau mass compared to the top one is due either to a small vev of $h_d$ (large $\tan\beta$) or a small component of $h_d$ in $\mathbf{16}_H$ (small $\theta_d$, see \eq{higgscomponents}). This can be seen from \eq{cpars} with $c_{D_3,L_3}\sim 1$, which gives $m_{b,\tau} = v\lambda_{b,\tau} \cos\beta \sim vh_{b,\tau}\sin\theta_d\cos\beta  \sim v\sin\theta_d\cos\beta $, where $v\approx 174\GeV$. From the point of view of the $A$-terms, the case with the third family of messengers close to the GUT scale is interesting because the contribution to the $A$-terms of fields with GUT scale masses is comparable to the one from the third family of messengers, and can significantly enhance them. For example, the $SU(5)$ representations $\mathbf{10}$ and $\mathbf{1}$ in the $\mathbf{16}$ or $\mathbf{16}'$ will contribute to the $A$-terms through their couplings to the matter fields in \eq{W2}. 

As mentioned, the contribution of the GUT-scale fields to the $A$-terms is quite model-dependent, as it depends on the detailed form of the $SO(10)$ lagrangian and on the implementation of doublet-triplet splitting in the Higgs sector. Still, a realistic estimate can be obtained by using the renormalizable part of the superpotential in \eq{W2} and by assuming that all the components in $\mathbf{16}$ and $\mathbf{16}'$ are at the same scale as the third family messengers. In such a case, the $A$-terms can be written as (neglecting corrections from non-renormalizable operators)
\globallabel{eq:trilinear104}
\begin{align}
A_U &=  A_{u^c} \lambda_U + \lambda_U A_q + \lambda_U A^{(u)}_y  \;, \mytag\\
A_D &=  A_{d^c} \lambda_D + \lambda_D A_q + \lambda_D (A^{(d)}_y + A_h + A_{h'}) \;, \mytag\\
A_E &=  A_{e^c} \lambda_E + \lambda_E A_l  + \lambda_E (A^{(d)}_y + A_h + A_{h'}) \;, \mytag
\end{align}
where the individual contributions read
\allowdisplaybreaks
\globallabel{eq:trilinearFull}
\begin{align}
A_q (M_{D_i}) &=  - \frac{1}{(4 \pi)^2} \frac{h^\prime_i}{h_i}  \left( 2 \left({h_i}^2 + {h^{\prime}_i}^2 \right) +  {y_i}^2 \right) \frac{F}{M} \;, \mytag\\  
A_{u^c} (M_{D_i}) &=  - \frac{1}{(4 \pi)^2} \frac{h^\prime_i}{h_i}   \left(  {h_i}^2 + {h^{\prime}_i}^2 +  2 {y_i}^2 \right)  \frac{F}{M} \;, \mytag\\  
A_{d^c} (M_{D_i}) &=  - \frac{1}{(4 \pi)^2} \frac{h^\prime_i}{h_i}   2 \left({h_i}^2 + {h^{\prime}_i}^2 \right)  \frac{F}{M} \;, \mytag\\
A_l (M_{D_i}) &= - \frac{1}{(4 \pi)^2} \frac{h^\prime_i}{h_i}  3 \left({h_i}^2 + {h^{\prime}_i}^2 \right)  \frac{F}{M} \;, \mytag\\
A_{e^c} (M_{D_i}) &= - \frac{1}{(4 \pi)^2} \frac{h^\prime_i}{h_i}   3 \left({h_i}^2 + {h^{\prime}_i}^2 \right)   \frac{F}{M} \;, \mytag\\
A^{(d)}_y (M_{D_i}) &= - \frac{3}{(4 \pi)^2} \frac{h^\prime_i}{h_i}  {y_i}^2\,   \frac{F}{M}  \;, \mytag\\[1mm]
A^{(u)}_y (M_{D_i}) &=  A_h (M_{D_i}) = A_{h'} (M_{D_i}) = 0 \;, \mytag\\
A_q (M_{L_i}) &=  - \frac{1}{(4 \pi)^2} \frac{h^\prime_i}{h_i}   \left(  {h_i}^2 + {h^{\prime}_i}^2  +  {y_i}^2 \right) \frac{F}{M}  \;, \mytag\\
A_{u^c} (M_{L_i}) &=  - \frac{1}{(4 \pi)^2} \frac{h^\prime_i}{h_i}   2 \left( {h_i}^2 + {h^{\prime}_i}^2  \right) \frac{F}{M}  \;, \mytag\\
A_{d^c} (M_{L_i}) &=  - \frac{1}{(4 \pi)^2} \frac{h^\prime_i}{h_i}  2 \left( {h_i}^2+ {h^{\prime}_i}^2  \right) \frac{F}{M}  \;, \mytag\\
A_l (M_{L_i}) &= - \frac{1}{(4 \pi)^2} \frac{h^\prime_i}{h_i}  \left({h_i}^2 + {h^{\prime}_i}^2  \right) \frac{F}{M}  \;, \mytag\\
A_{e^c} (M_{L_i}) &= - \frac{1}{(4 \pi)^2} \frac{h^\prime_i}{h_i} \,  2 {y_i}^2 \frac{F}{M}   \;, \mytag\\
A^{(d)}_y (M_{L_i}) &= - \frac{1}{(4 \pi)^2} \frac{h^\prime_i}{h_i}    {y_i}^2  \frac{F}{M}   \;, \mytag\\
A^{(u)}_y (M_{L_i}) &= - \frac{1}{(4 \pi)^2} \frac{h^\prime_i}{h_i}    {y_i}^2  \frac{F}{M}   \;, \mytag\\
A_h (M_{L_i}) &= - \frac{1}{(4 \pi)^2} \frac{h^\prime_i}{h_i}    {h_i}^2  \frac{F}{M}   \;, \mytag\\
A_{h'} (M_{L_i}) &= - \frac{1}{(4 \pi)^2} \frac{h^\prime_i}{h_i}  {h^{\prime}_i}^2  \frac{F}{M}   \;. \mytag
\end{align}
In our numerical analysis we will use for definiteness the above expressions.

\subsection{Two loop level contributions to sfermion masses}
\label{sec:sfermions}

The coupling of the chiral messengers $D^c$, $\overline{D^c}$, $L$ and $\overline{L}$ to SUSY breaking, eq.~\eqref{eq:WZ}, gives rise to the well known MGM two loop contributions to sfermion masses. In this section we give their expressions in our model. As the chiral messengers have supersymmetric masses $h_{D_i} M$ and $h_{L_i} M$ and supersymmetry breaking mass terms given by ${h_D^\prime}_i F$ and ${h_L^\prime}_i F$, the contributions to sfermion masses, as the ones to gaugino masses, depend on the parameters ${\gamma_D}_i$ and ${\gamma_L}_i$ and can be similarly ehanced. We have in fact 
\begin{equation}
\begin{split}
(m^2_Q)_{\text{MGM}} &= \sum_i (m^2_Q)_{\text{MGM}} ({M_D}_i) + (m^2_Q)_{\text{MGM}} ({M_L}_i) \\
&= 2 \bigg[ \bigg(c^{(3)}_Q  \frac{\alpha_3^2({M_D}_i)}{(4 \pi)^2} + \frac{2}{5} c^{(1)}_Q  \frac{\alpha_1^2({M_D}_i)}{(4 \pi)^2} \bigg) {\gamma_D}_i^2  \; \\ & \quad\,+ \bigg(c^{(2)}_Q  \frac{\alpha_2^2({M_L}_i)}{(4 \pi)^2} + \frac{3}{5} c^{(1)}_Q  \frac{\alpha_1^2({M_L}_i)}{(4 \pi)^2} \bigg) {\gamma_L}_i^2  \bigg] \left( \frac{F}{M}\right)^2 \; , \\ 
\end{split}
\end{equation}
where $c^{(a)}_Q$ is the quadratic Casimir of the sfermion $\widetilde{Q}$ (or Higgs $Q$) relative to the gauge interaction $a$, as in Table \ref{chap:TGM_LHC_tab:Casimirs}. The parameters $\gamma_{D_i,L_i}(F/M)$ are determined by the parameters $M_{1/2}$, $r$, ${r_D}_i,{r_L}_i$, $i=1,2$ through eqs.~\eqref{eq:rLD}, \eqref{eq:gamma} and \eqref{eq:M12}. 
\begin{table}
\centering
\begin{tabular}{cccccccc}
\toprule
$Q$ & $q_i$ & $u_i^c$ & $d_i^c$  & $l_i$ & $e_i^c$ & $h_u$ & $h_d$ \\ 
\midrule
$c^{(1)}_Q$ & $1/60$ & $4/15$ & $1/15$ & $3/20$ & $3/5$ & $3/20$ & $3/20$  \\[0.3pc]
$c^{(2)}_Q$ & $3/4$ & $0$ & $0$ & $3/4$ & $0$ & $3/4$ & $3/4$  \\[0.3pc]
$c^{(3)}_Q$ & $4/3$ & $4/3$ & $4/3$ & $0$ & $0$ & $0$ & $0$  \\
\bottomrule
\end{tabular}
\caption{Quadratic Casimirs for the low energy superfields. \label{chap:TGM_LHC_tab:Casimirs}}
\end{table}

On top of the usual MGM contributions, soft masses receive also two loop contributions because of messenger-matter mixing. Sizeable contributions arise only for third family sfermions (and Higgses). All in all the corrections are (remember that $|F/M|^2 = 10 \, m_{10}^2$)
\globallabel{1021}
\begin{align}
(4 \pi)^4 \delta {m}^2_{q_3} =&  \left( \frac{7}{30} g_1^2  + \frac{3}{2}  g_2^2  + \frac{8}{3} g_3^2 - 3  \hat{\lambda}_{D_3}^2 - \frac{1}{2}( \lambda_{E_3}^2 + \hat{\lambda}_{E_3}^2) \right) \hat{\lambda}_{D_3}^2 \gamma_{D_3}^2 \left( \frac{F}{M} \right)^2 \nonumber \\
&+ \frac{1}{2} \lambda_{D_3}^2 \hat{\lambda}_{E_3}^2 \gamma_{E_3}^2 \left( \frac{F}{M} \right)^2  \mytag
\\
(4 \pi)^4 \delta {m}^2_{l_3} =& \left( \frac{3}{2}  \hat{\lambda}_{D_3}^2 \lambda_{E_3}^2 \gamma_{D_3}^2 + 2  \lambda_{E_3}^2 \hat{\lambda}_{E_3}^2 \gamma_{E_3}^2 \right) \left( \frac{F}{M} \right)^2  \mytag
\\
(4 \pi)^4 \delta {m}^2_d =&  \left(  6 \hat{\lambda}_{D_3}^2 \lambda_{D_3}^2 \gamma_{D_3}^2 + \hat{\lambda}_{E_3}^2 \lambda_{D_3}^2 \gamma_{E_3}^2 \right) \left( \frac{F}{M} \right)^2  \mytag
\\
(4 \pi)^4 \delta {m}^2_{e_3}  =&  \left( \frac{9}{5} g_1^2 +3 g_2^2 - 4  \hat{\lambda}_{E_3}^2 - 3 ( \lambda_{D_3}^2 + \hat{\lambda}_{D_3}^2) \right) \hat{\lambda}_{E_3}^2 \gamma_{E_3}^2 \left( \frac{F}{M} \right)^2  \nonumber \\
& + 3 \lambda_{E_3}^2 \hat{\lambda}_{D_3}^2 \gamma_{D_3}^2 \left( \frac{F}{M} \right)^2  \mytag
\\
(4 \pi)^4 \delta {m}^2_{u_3} =&  \left( \lambda_{U_3}^2 \hat{\lambda}_{D_3}^2 \gamma_{D_3}^2  \right) \left( \frac{F}{M} \right)^2  \mytag
\\
(4 \pi)^4 \delta {m}^2_{h_d} =&  \left( \frac{7}{10} g_1^2 + \frac{9}{2} g_2^2 + 8 g_3^2 - 9  \hat{\lambda}_{D_3}^2 - \frac{3}{2} (\hat{\lambda}_{E_3}^2 + \lambda_{U_3}^2 ) \right)\hat{\lambda}_{D_3}^2\gamma_{D_3}^2 \left( \frac{F}{M} \right)^2 \nonumber \\
&+ \left( \frac{9}{10} g_1^2 + \frac{3}{2} g_2^2  - 2  \hat{\lambda}_{E_3}^2 - \frac{3}{2}  \hat{\lambda}_{D_3}^2 \right) \hat{\lambda}_{E_3}^2 \gamma_{E_3}^2 \left( \frac{F}{M} \right)^2     \mytag
\\
(4 \pi)^4 \delta {m}^2_{h_u} =&  \frac{3}{2}  \hat{\lambda}_{D_3}^2 \lambda_{U_3}^2 \gamma_{D_3}^2  \left( \frac{F}{M} \right)^2   \mytag \, .
\end{align}

\section{Analysis of the parameter space}
\label{sec:analysis}

Let us now discuss the parameter space of the model. As pointed out in section~\ref{sec:RelevantParameters}, the relevant parameters to be specified are $m_{10}$, $M_{1/2}$, $r$, $\tan\beta$, $\text{sign}(\mu)$, $\theta_u$ and $\theta_d$. 
Let us begin from a discussion of the allowed range for the angles $\theta_u$, and $\theta_d$. 

\subsection[Allowed ranges of $\theta_u$ and $\theta_d$]{Allowed ranges of $\boldsymbol{\theta_u}$ and $\boldsymbol{\theta_d}$}

Two constraints have to be taken into account: reproducing the SM fermion masses and EWSB. Since the top Yukawa coupling is essentially given by $\lambda_t = y_{3} \cos\theta_u$, see appendix~\ref{sec:parameters}, we should have $\cos\theta_u = \ord{1}$, if $y_3$ has to be kept perturbative and possibly of order one, as $\lambda_t$. Which means that $\cos\theta_u$ should be sizeable, with the maximal value $\cos\theta_u=1$ also allowed. Similarly, as the bottom Yukawa coupling is given by $\sin\theta_d$ times a combination of couplings  that we expect not to be much larger than 1 (see appendix~\ref{sec:parameters}), we should have $\sin\theta_d \gtrsim \lambda_b  = m_b/(\cos\beta v) \sim 10^{-2}\tan\beta$.  In summary we have
\begin{equation}
\label{eq:thetabound1}
\begin{aligned}
\cos\theta_u &\sim \ord{1} \\ \sin\theta_d &\gtrsim 10^{-2} \tan\beta
\end{aligned}
\end{equation}
from the requirement of perturbativity of the couplings generating the SM fermion masses. 

\begin{figure}
 \centering
\includegraphics[width=0.50\textwidth]{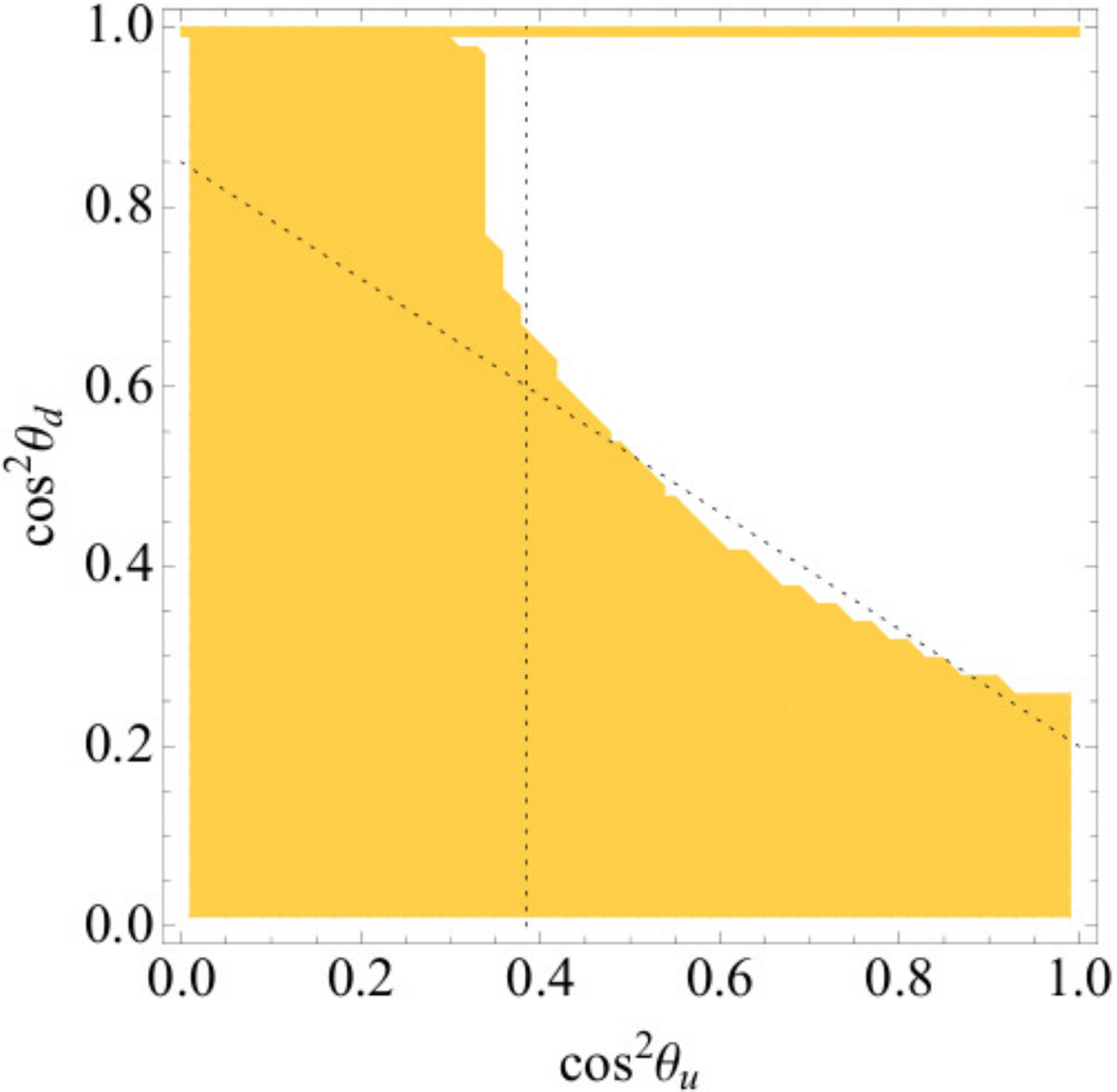}
 \caption{Constraints on $\theta_u$ and $\theta_d$ from proper breaking of the EW symmetry and perturbativity. The figure has been obtained for $\tan\beta=10$, $m_{10} = 1.8\TeV$, $M_{1/2} = 600\GeV$, $r = 1$. Also shown are the approximate constraints in eq.~\eqref{eq:thetabound2} (dotted lines).}
\label{fig:cosines}
\end{figure}

The angles $\theta_u$ and $\theta_d$ also enter the EWSB conditions through the tree level expression for the Higgs soft masses. In order for EWSB to take place for a given value of $\tan\beta$ (and $M_Z$), the following two conditions have to be satisfied: 
\begin{equation}
\label{eq:ewsbconditions}
\begin{gathered}
\frac{m^2_{h_d} - m^2_{h_u}\tan^2\beta}{\tan^2\beta -1} \geq M^2_Z /2 \\ 
(m^2_{h_d} - m^2_{h_u})\frac{\tan^2\beta + 1}{\tan^2\beta - 1} + M^2_Z > 0 \,.
\end{gathered}
\end{equation}
For moderately large values of $\tan\beta$ and in the typical fine tuned situation in which $|m^2_{h_u}| \gg M^2_Z$, the latter conditions become $m^2_{h_u} \lesssim 0$ and $m^2_{h_d} - m^2_{h_u} \gtrsim 0$. The corresponding constraints on $\theta_u$ and $\theta_d$ can be obtained in analytical form in the limit in which eqs.~\eqref{eq:running} hold (a typical fine tuned scenario with moderately large $\tan\beta$ and sfermions heavier than gauginos): 
\begin{equation}
\label{eq:thetabound2}
\cos^2\theta_d + \Big(1-\frac{\rho}{2}\Big) \cos^2\theta_u \gtrsim \frac{6}{5}-\frac{\rho}{2} \qquad
\cos^2\theta_u \gtrsim \frac{3/5-\rho/2}{1-\rho/2} \,.
\end{equation}
Finally, some values of $\cos\theta_u$ and $\cos\theta_d$ may not be allowed even if the constraints in eq.~\eqref{eq:ewsbconditions} hold, for example because some particle becomes tachyonic.

The constraints on $\theta_u$ and $\theta_d$ from proper EWSB should be merged with the ones from fermion masses (eqs.~\eqref{eq:thetabound1}). The constraint $\theta_u = \ord{1}$ is automatically satisfied once eqs.~\eqref{eq:ewsbconditions} hold, while the constraint on $\theta_d$ in eqs.~\eqref{eq:thetabound1} cuts an additional thin stripe of parameter space close to the $\cos^2\theta_d = 1$ axis. The overall constraint one gets in the $\cos^2\theta_u$--$\cos^2\theta_d$ plane is shown (for fixed values of the other parameters) in figure~\ref{fig:cosines}. The allowed points with $\cos^2\theta_u$ near the left vertical bound (where $m^2_{h_u}$ changes sign) correspond to smaller $|m^2_{h_u}|$ and therefore relatively smaller fine-tuning. We see from the figure that a pure embedding of the MSSM up Higgs in the $\mathbf{10}_H$ (with no component in $\overline{\mathbf{16}}_H$, $\cos\theta_u = 1$) is allowed, while the down Higgs must have a mixed embedding, with components in both the $\mathbf{16}_H$ and $\mathbf{10}_H$. A component in the $\mathbf{16}_H$ is needed to obtain non vanishing down quark masses (at the tree, renormalizable level), while a component in the $\mathbf{10}_H$ is necessary for a correct EWSB.

\subsection{A 125$\,$GeV Higgs}

\begin{figure}
 \centering
 \includegraphics[scale=0.7]{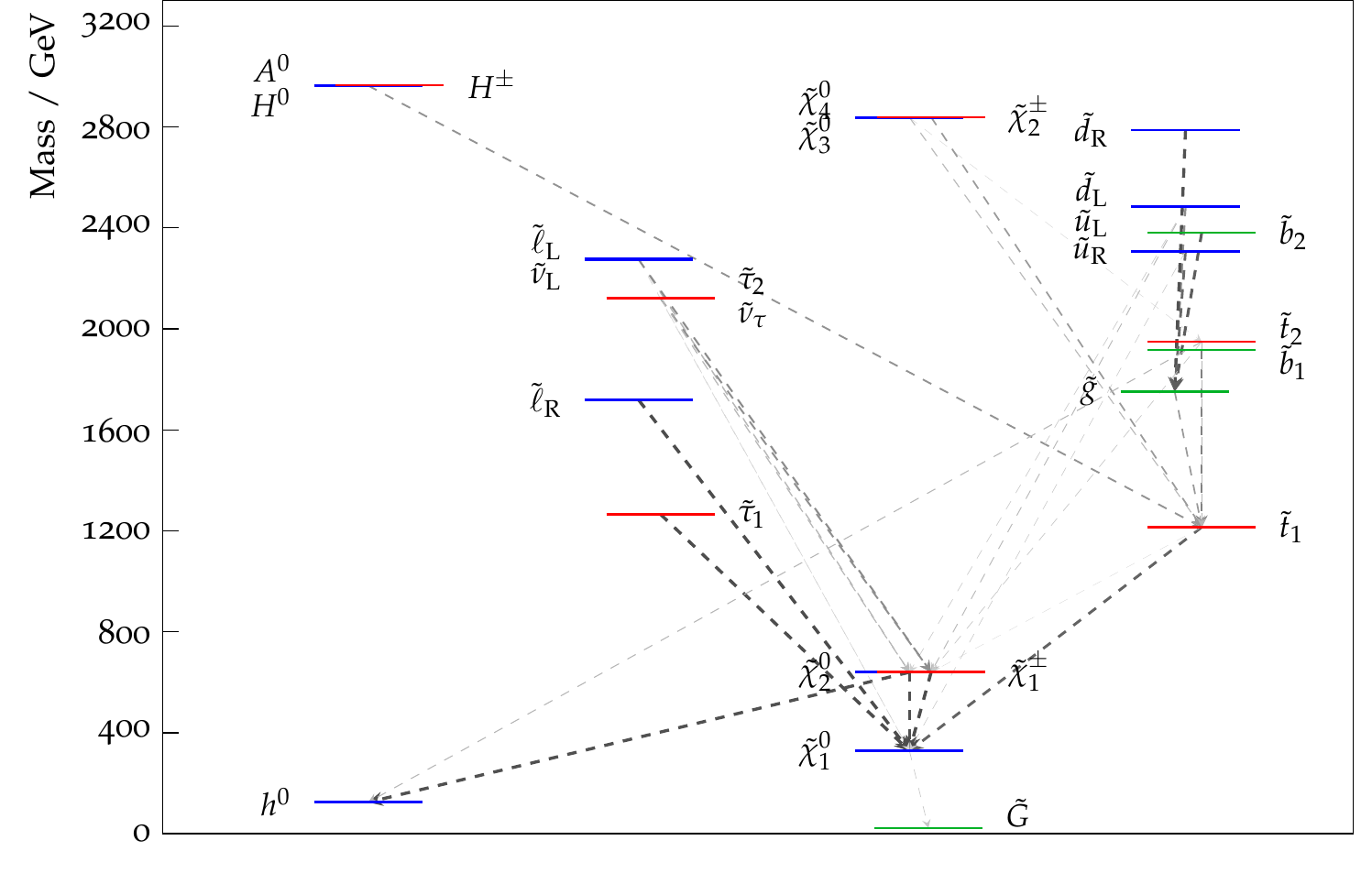}
 \caption{The SUSY spectrum of a point with a Higgs mass of 125~GeV, calculated with our modified version of the {\tt softSUSY} package \cite{Allanach:2001kg}. The decays are depicted by the dashed grey arrows, which are scaled with the respective BR calculated via {\tt SUSY-HIT} \cite{Djouadi:2006bz}: only BRs greater than 0.1 are shown.  The point corresponds to  $m_{10} = 1.5~\TeV$, $\cos \theta_u = 0.9$, $\cos \theta_d = 0.9$, $\gamma = 16.7$, $r = 1$, ${r_D}_{1,2} = {r_L}_{1,2} = 0.3$, $\tan \beta = 10$, $\text{sign}(\mu) = \mathbf{+}$. \label{fig:SpectrumHiggs}}.
\end{figure}

In standard gauge mediation it is not easy to accommodate a rather heavy Higgs boson with a mass of about $125\GeV$, as indicated by the recent evidence \cite{higgsdiscovery1,higgsdiscovery2}. Such a mass needs in fact moderately large $\tan \beta$ and a rather heavy SUSY scale or large trilinear couplings, see, e.g.\ \cite{Arbey:2011ab}. In standard gauge mediation it is usually assumed that the messengers have only gauge interactions with the SM fields and hence the trilinear couplings are strongly suppressed at the messenger scale. RGE running does  give rise to a non-negligible contribution to the $A$-terms, but not large enough~\cite{Draper:2011aa}. Sizeable trilinear terms can be generated by introducing superpotential messenger matter interactions. However, the latter potentially spoil the flavour universality of the soft terms, one of the main motivations for gauge mediation models (see however~\cite{MYGM,Evans:2011bea1,Evans:2011bea2}). 

Things are different in our setup. Sizeable trilinears  are generated because the messengers unavoidably have Yukawa couplings to the MSSM fields, as we discussed in section~\ref{sec:Aterms}. Such trilinears arise at the one loop level but they turn out to enjoy a potential enhancement by the same parameter $\gamma$ enhancing gaugino masses. Moreover, because of the $SO(10)$ relations between them, the flavour structure of the messenger matter couplings is dictacted by the SM Yukawas. As a consequence, they do not spoil the solution of the supersymmetric flavour problem offered by our framework.
A spectrum reproducing a light Higgs of 125~GeV is shown in figure \ref{fig:SpectrumHiggs}.

Alternatively the Higgs mass can be increased above the MSSM values in the presence of a mixing with a SM singlet chiral field $S$, as in the NMSSM~\cite{Ellwanger:2009dp}. In MGM, such a SM singlet would have vanishing soft mass at the messenger scale, as it does not couple to SM gauge interactions. This is not necessarily the case in TGM, as the soft masses are generated by $U(1)_X$ gauge interactions. Depending on the $SO(10)$ embedding of the $Sh_u h_d$ interaction lifting the Higgs mass, such a singlet could acquire a positive, vanishing, or negative soft mass. In fact, let us remind that $h_u$ can be embedded into a $\mathbf{10}$ or a $\overline{\mathbf{16}}$, while $h_d$ can be embedded into a $\mathbf{10}$ or a $\mathbf{16}$. We therefore have 4 possibilities for the $SO(10)$ embedding of the $Sh_u h_d$ interaction: $\mathbf{1}_S \mathbf{10}_{h_u} \mathbf{10}_{h_d}$, $\overline{\mathbf{16}}_S \overline{\mathbf{16}}_{h_u} \mathbf{10}_{h_d}$, $\mathbf{16}_S \mathbf{10}_{h_u}\mathbf{16}_{h_d}$, $\mathbf{1}_S \overline{\mathbf{16}}_{h_u} \mathbf{16}_{h_d}$ (where $\mathbf{1}_S$ can be substituted by $\mathbf{45}_S$ or $\mathbf{54}_S$ without affecting our conclusions). The soft terms of the singlet $S$ is correspondingly given by $\tilde m^2_S = 0, -5m^2_{10}, 5 m^2_{10}, 0$. If the soft mass is negative, a vev for the $S$ field (and a solution for the $\mu$ problem) can be induced. In the following, we will take into account the possibility of an NMSSM-like extra contribution to the Higgs mass. However, we will not enter the model building details associated to the possible presence of a NMSSM singlet in the TeV scale spectrum, leaving them to forthcoming studies.

\subsection{NLSP}

\begin{figure}
 \centering
 \includegraphics[scale=0.50]{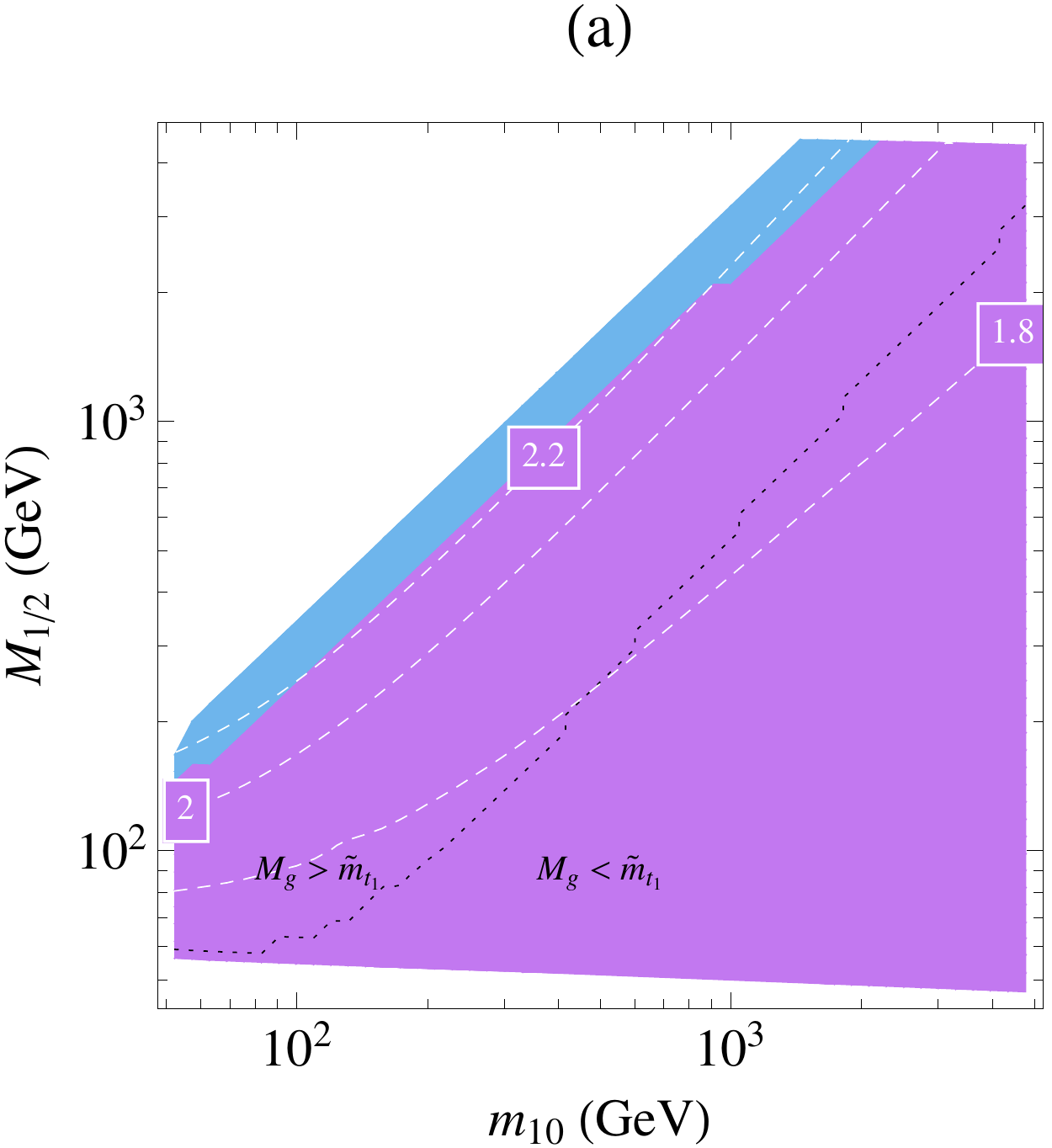} \hfill
 \includegraphics[scale=0.50]{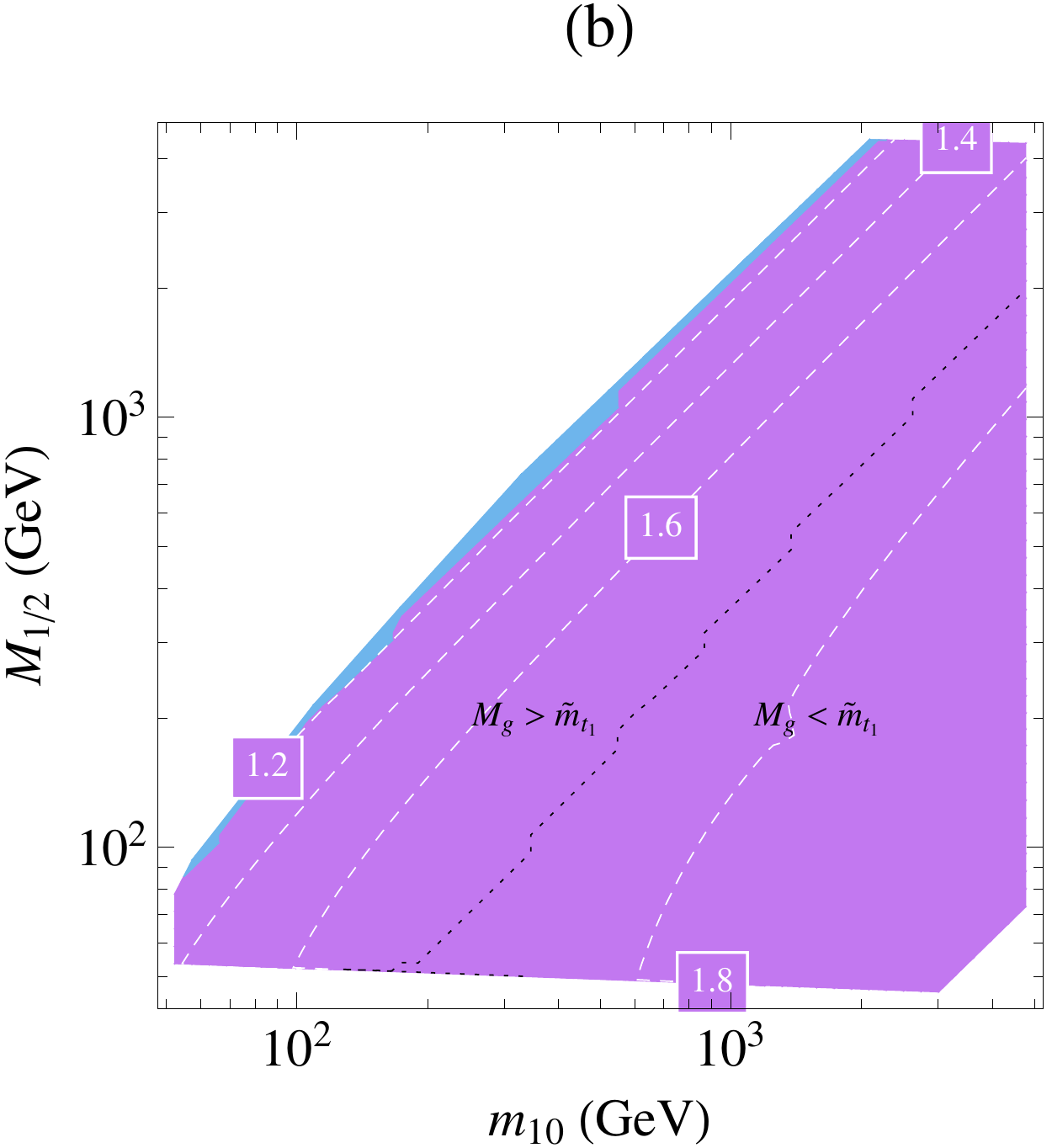} 
 \caption{Nature of the NLSP in the allowed $m_{10}$--$M_{1/2}$ parameter space for  $\tan\beta=10$, $\cos\theta_u = 0.9$, $\cos\theta_d = 0.9$, $\text{sign}(\mu) = 1$, and $r = 1$ (a) or $r = 0.2$ (b). The NLSP is a neutralino in the violet region and a stau in the light blue region. The violet region corresponds to a bino like neutralino in the left panel (a) and to a wino like neutralino in the right panel (b). The regions in which the lightest coloured particle is a stop or a gluino are separated by a black dotted line. Also shown is the ratio $m^2_{\tilde e_L}/m^2_{\tilde e_R}$ of left and right handed squared selectron masses (white yellow lines). 
 \label{fig:nlsp} }
\end{figure}

In TGM models, the Lightest Supersymmetric Particle (LSP) is the gravitino. The cosmology of the model is therefore determined first of all by the nature of the Next to LSP (NLSP) which has a lifetime of hundreds of picoseconds in our benchmark points. The cosmology of such a NLSP is a interesting subject on its own and was studied before elsewhere~\cite{TGM1,Arcadi:2011yw} so that we will not discuss it here further.
The NLSP turns out to be a neutralino or the stau, depending on the region of the parameter space. Whether the lightest neutralino is bino like or wino like is essentially determined by the parameter $r$, as illustrated by figure \ref{fig:gauginos}. When $r \gtrsim 0.3$, the NLSP is either a bino like neutralino or a stau, while when $r \lesssim 0.3$ the NLSP is either a wino like neutralino or a stau. Figure \ref{fig:nlsp} shows the part of the parameter space in which the NLSP is a neutralino (violet) or a stau (light blue). On the left panel, $r = 1$ and the neutralino is bino like, while on the right panel $r = 0.2$ and the neutralino is wino like. The remaining parameters are $\tan\beta=10$, $\cos\theta_u = 0.9$, $\cos\theta_d = 0.9$, $\text{sign}(\mu) = 1$.
The figure shows that the NLSP is a neutralino in most of the parameter space. On the other hand, a stau stripe is present in both cases. This is because the upper left boundary of the parameter space is due to the stau becoming tachyonic. A stau NLSP can therefore be obtained in a region close enough to that boundary. The regions in which the lightest coloured particle is the lightest stop or the gluino are separated by a dotted line. Finally, the ratio of left and right handed squared selectron masses is also shown (dashed white lines). As a peculiar prediction of the minimal $SO(10)$ TGM scenario, that ratio is predicted to be two at the tree level. A calculable deviation from two is induced by loop corrections due to RGE running and minimal gauge mediation effects. The figure shows  the prediction for the $m^2_{\tilde e_L}/m^2_{\tilde e_R}$ mass ratio, including the radiative correction. In the Bino NLSP case, the radiative corrections have a smaller impact (up to 10\%) on the tree-level value, while in the Wino case, the impact can reach 20--30\%. 

In the light of the discussion above, we will consider three representative points in the parameter space in which the NLSP is a bino like neutralino, a wino like neutralino or a stau. 

\subsection{Three benchmark points}
\label{chap:TGM_LHC_benchs}

TGM models can provide a variety of signatures at the LHC.  The nature
of the NLSP and its long lifetime dictate the phenomenology. When the
neutralino is the NLSP, a classical CMSSM-like phenomenology: colored
sparticles are produced in the collision. The subsequent cascade
generates events with missing transverse energy, jets, and possibly
leptons. The decay of the NLSP to gravitino happens outside the
detector. 

When the NLSP is a charged particle (e.g. staus), SUSY could
be found looking for heavy stable charged particles (HSCP). This kind
of signature usually implies a dedicated reconstruction of the HSCP,
which crosses the detector layers out of time with respect to the
other particles (being slower). One then needs to connect different
hits in different bunch crossings. At the same time, just looking at
the collision bunch crossing (as it is done in the standard
reconstruction) one typically fails to reconstruct the HSCP. The rest
of the SUSY event will then look like a typical event with MET, as in
the case of neutralino NLSP. 

The phenomenology changes whenever the squarks and gluons are above
the TeV. The squark-squark cross section becomes negligible for the
luminosity collected by LHC for the first run. The main production
mechanisms are gluino-squark and gluino-gluino. This implies that,
despite the 3rd generation squarks being the lightest, their
production is not dominant. The production of charginos and
neutralinos (ewkinos) is suppressed by the coupling but enhanced by
the low mass and it could become the dominant production
mechanism. The detection of these events is challenging for the LHC
experiments, when the ewkinos are close in mass and only soft
particles are produced in the decay. The SUSY production with
associated jets is then the most effective process to access these
events, for instance with a monojet or a dijet analysis.

We consider three benchmark points with different NLSP, to highlight
the main phenomenological implications with specific examples. Let us
discuss their main features before entering the details of collider
searches.

\begin{table}
\centering
\begin{tabular}{lccccccccc}
\toprule
Point & $m_{10}$ in TeV & $\cos \theta_u$ & $\cos \theta_d$ & $\gamma$ & $r$ & $\tan \beta$ & $\text{sign }(\mu)$ \\ \midrule
Bino & 1.0 & 0.9 & 0.9 & 15 & 1 & 10 & +1 \\
Wino & 0.55 & 0.9 & 0.9 & 20 & 0.2 & 10 & +1 \\
Stau & 0.8 & 0.9 & 0.9 & 35 & 1 & 35 & +1 \\
\bottomrule
\end{tabular}
\caption{TGM parameters for our three benchmark points with the NLSP as specified and for the point with the light Higgs mass of about 125~GeV.
\label{tab:BenchPoints}}
\end{table}

\subsubsection{Bino NLSP benchmark point}

\begin{figure}
\centering
\includegraphics[scale=0.75]{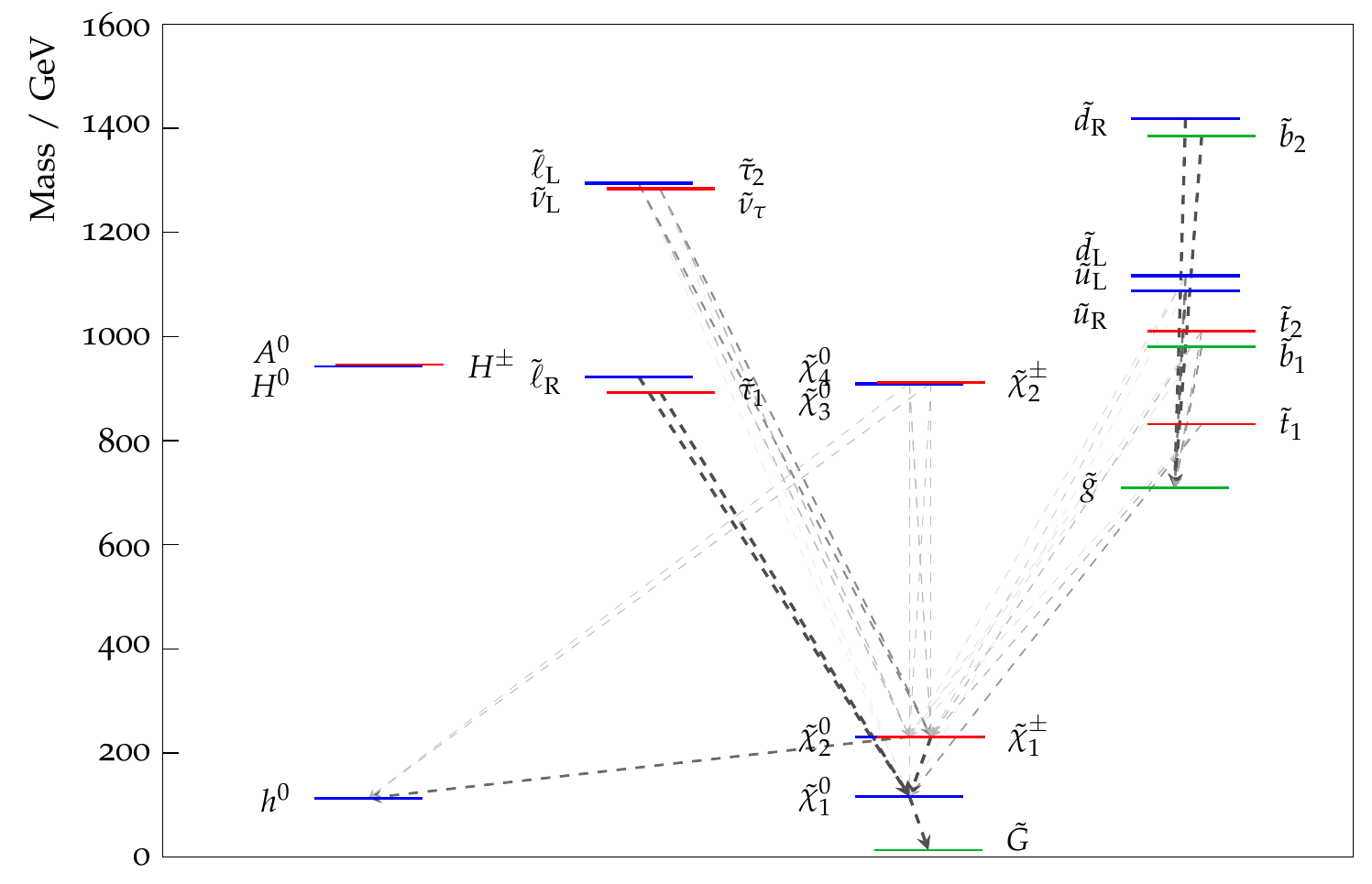}
\caption{The SUSY spectrum of the benchmark point with bino NLSP calculated with our modified version of the {\tt softSUSY} package \cite{Allanach:2001kg}. The decays are depicted by the dashed grey arrows, which are scaled with the respective BR calculated via {\tt SUSY-HIT} \cite{Djouadi:2006bz}: only BRs greater than 0.1 are shown.  The point corresponds to  $m_{10} = 1.0~\TeV$, $\cos \theta_u = 0.9$, $\cos \theta_d = 0.9$, $\gamma = 15$, $r = 1$, $\tan \beta = 10$, $\text{sign}(\mu) = \mathbf{+}$. \label{fig:SpectrumBino}}
\end{figure}

The case in which the NLSP is a bino like neutralino is the most common one if $r$ is not too small. As figure \ref{fig:nlsp} shows, the $m^2_{\tilde e_L}/m^2_{\tilde e_R}$ ratio is typically within 10\% of the tree-level prediction, even for a relatively light spectrum. 

In the case of the benchmark point we choose, corresponding to the spectrum in figure \ref{fig:SpectrumBino}, the typical final states at the LHC are characterized by a large presence of $b$-enriched final states accompanied by multileptonic signals. The $b$ quarks and leptons largely come from the electroweak decays of the charginos and neutralinos
down to the NLSP. The gaugino mass separation allows the interesting possibility that the lightest Higgs is produced in cascade decays, as the $\chi_2^0 \to \chi_1^0 H$ decay is kinematically allowed, in turn characterized by the subsequent on-shell decay to $b$ quark pairs.
Because of the large MET associated to the NLSP, which escapes detection before decaying to the gravitino, the characteristic feature of such models would be the presence of both SUSY signatures and the Higgs boson in the same event. The latter situation makes it profitable to consider such a scenario both with inclusive and exclusive dedicated searches as we shall see in the following. 

\subsubsection{Wino NLSP benchmark point}

\begin{figure}
\centering
\includegraphics[scale=0.75]{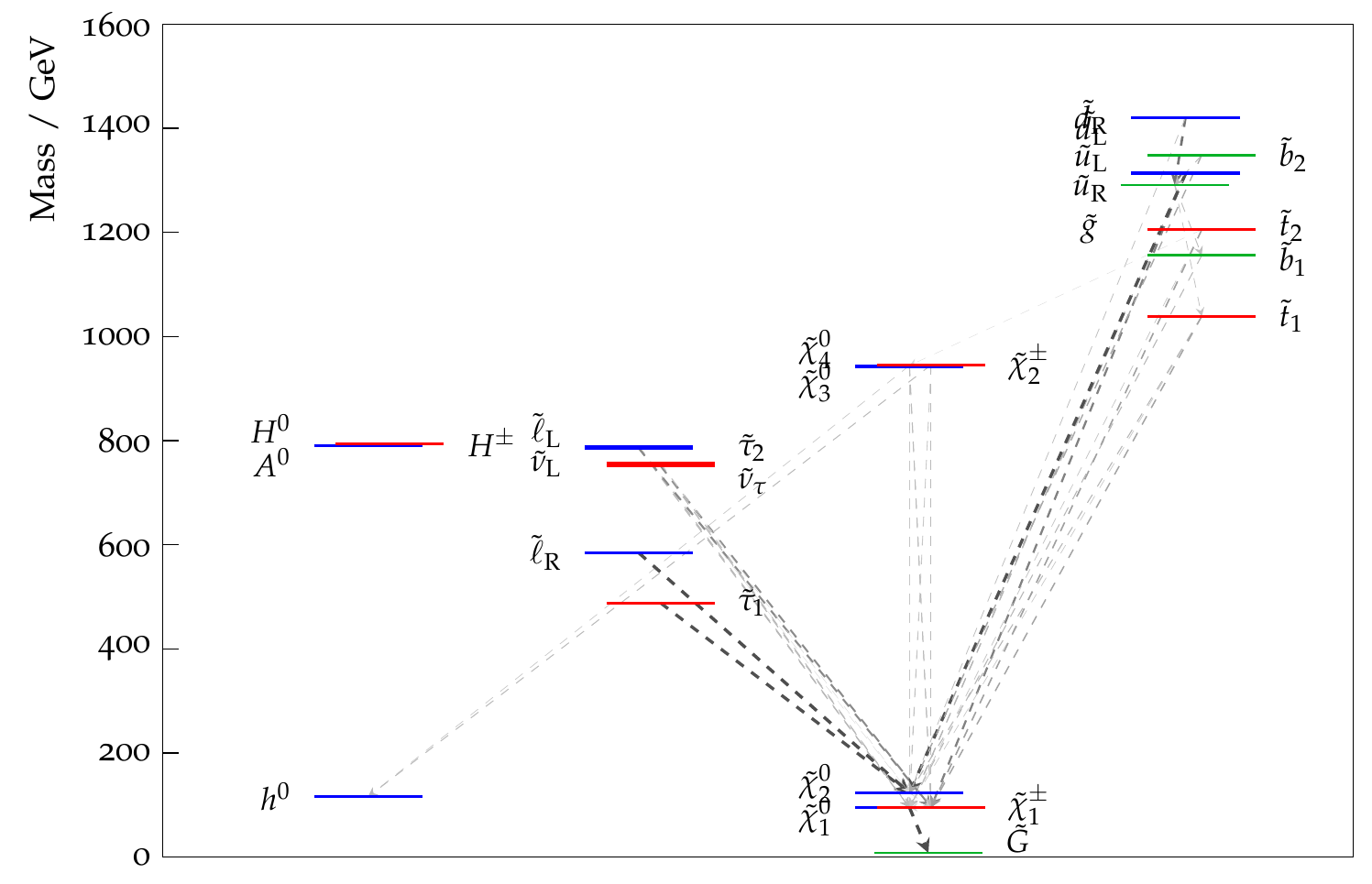}
\caption{The SUSY spectrum of the benchmark point with wino NLSP calculated with our modified version of the {\tt softSUSY} package \cite{Allanach:2001kg}. The decays are depicted by the dashed grey arrows, which are scaled with the respective BR calculated via {\tt SUSY-HIT} \cite{Djouadi:2006bz}: only BRs greater than 0.1 are shown.  The point corresponds to  $m_{10} = 550~\GeV$, $\cos \theta_u = 0.9$, $\cos \theta_d = 0.9$, $\gamma = 20$, $r = 0.2$, $\tan \beta = 10$, $\text{sign}(\mu) = \mathbf{+}$.  \label{fig:SpectrumWino} }
\end{figure}

The case in which the NLSP is a wino like neutralino usually leads to a heavier spectrum than obtained in the bino case. The tree-level prediction $\tilde m^2_{l} = 2 \tilde m^2_{e^c}$, $\tilde m^2_{d^c} = 2 \tilde m^2_{q,u^c}$ gives rise to a separation between two groups of soft masses in the light families of both the slepton and squark sector. The inverted hierarchy between the two lightest gaugino masses, $M_2 < M_1$, makes the lightest chargino and the lightest neutralino approximately degenerate, as they have both mass $M_2$ before EWSB. This makes the decay into the NLSP particularly soft and makes the decay $\chi_2^0 \to \chi_1^0 H$ kinematically forbidden, unlike what discussed in the bino NLSP case. From this point of view, it is then comparatively more profitable to use semi- and full-leptonic channels, because of the absence of $H \rightarrow b \overline{b}$ in the decay chain.

\subsubsection{Stau NLSP benchmark point}
\label{chap:TGM_LHC_benchs_stau}

\begin{figure}
\centering
\includegraphics[scale=0.75]{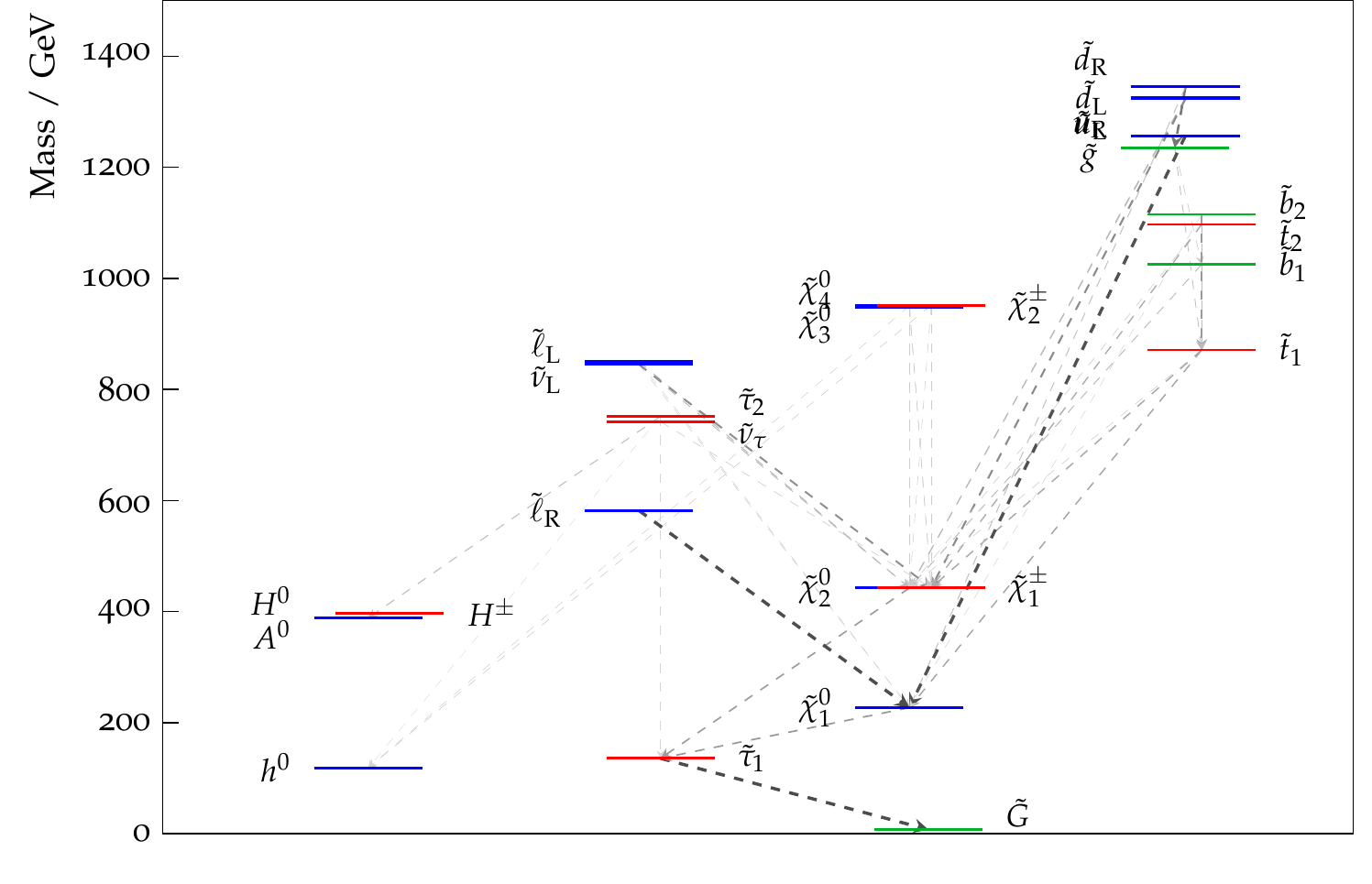}
\caption{The SUSY spectrum of the benchmark point with stau NLSP calculated with our modified version of the {\tt softSUSY} package \cite{Allanach:2001kg}. The decays are depicted by the dashed grey arrows, which are scaled with the respective BR calculated via {\tt SUSY-HIT} \cite{Djouadi:2006bz}: only BRs greater than 0.1 are shown.  The point corresponds to  $m_{10} = 800~\GeV$, $\cos \theta_u = 0.9$, $\cos \theta_d = 0.9$, $\gamma = 35$, $r = 1$, $\tan \beta = 35$, $\text{sign}(\mu) = \mathbf{+}$.  \label{fig:SpectrumStau} }
\end{figure}

The gaugino masses are determined by the parameter $M_{1/2}$, while the sfermion masses (at the tree level) are associated to $m_{10}$. For larger $M_{1/2}/m_{10}$, one therefore expects the NLSP to be the lightest sfermion, i.e.\ the lightest stau. This is the case, but only in a small portion of the parameter space, as the radiative corrections to the stau mass proportional to the gaugino masses can make the stau leptons heavier than the gauginos even for largish $M_{1/2}/m_{10}$. The stau NLSP region in figure~\ref{fig:nlsp} are close to the upper-left border of the parameter space, which is associated to the stau becoming more than light: tachyonic. In the regions of parameter space characterized by a stau NLSP the tree level and the $M_{1/2}$-driven radiative contributions to coloured sfermion masses are comparable. It is therefore necessary to include the latter contribution in order to test the TGM prediction for the sfermion mass ratios. The stau  is not expected to decay to the gravitino in the detector. One can then use searches for heavy charged stable particles, on top of inclusive ones.

\section{TGM phenomenology at the LHC}
\label{sec:razor}

The search for SUSY with MET at the LHC has made remarkable progresses
with respect to the previous experiments. The favourable beam energy
and the large luminosity collected are the basic ingredients that
determined this improvement. On the other hand, many progresses have
been made also on the analysis technique, with new ideas introduced to
suppress the background and increase the signal sensitivity.  The
ATLAS and CMS experiments have collected so far $\sim~5$ fb$^{-1}$ at
7 TeV and are expected to collect $\sim~20$ fb$^{-1}$ at 8 TeV. The
current limits are pushing the masses of the coloured superpartners
above the 1 TeV threshold for generic MSSM models
\cite{squarksATLAS,squarksCMS}, while lower masses are allowed for
stop and sbottom in the case of models with large mass splitting among
the third family and the others \cite{stopATLAS,razor}. So far, the
possibility of light charginos and neutralinos has been tested only
through multi lepton final states \cite{charginoATLAS,charginoCMS},
which suffer from the suppression coming from $Z\to \ell \ell$ and
$W\to \ell \nu$ branching ratios. The increase in the center of mass
energy will be beneficial to push the mass limits on squarks and
gluino above the TeV scale, while the search for light EW gauginos
will be pushed by the larger collected luminosity.

In this scenario, a possible hint of new physics could emerge by the
end of 2012, but even in this situation the mission would be far from
being accomplished. The search for SUSY would be completed by the
characterization of a possible excess in terms of a specific SUSY
model, to possibly underline the nature of the SUSY breaking mechanism
and of its mediation. Accomplishing this goal, sometimes referred to
as the {\it inverse LHC problem} \cite{ArkaniHamed:2005px}, would
imply the use of kinematic variables sensitive to the mass of the
produced particles in as many final states as possible.

The TGM class of models offers a rich phenomenology at the LHC,
challenging the experiments on several fronts at the same time
(e.g. high mass searches, compressed gaugino spectra, $\dots$) and
allowing several many interesting possibilities, such as Higgs
production in SUSY cascades. In this respect, TGM is an interesting
playground on which the performances of different searches
(e.g. hadronic vs. leptonic searches) could be compared, and, on top
of that, it comes with a specific prediction on the ratio of sfermion
masses, which should be tested by experiments in case an excess is
found.

A full review of all the analyses presented by ATLAS and CMS and their
implications on TGM goes beyond the scope of this paper.  Instead, we
consider only the CMS razor analysis~\cite{razorpaper, razor}, which
offers a set of interesting features:
\begin{enumerate}
\item It considers simultaenously
six final states ($1\mu 1e$, $2 \mu$, $2 e$, $1 \mu$, $1 e$, and
hadronic) providing in one goal the combination of six different
analyses.
\item It gives a competitive limit on all the signatures it is
sensitive to, giving a reasonable estimate of the current constraints
from the full LHC SUSY program.
\item Besides being sensitive to a signal, it also offers some
  information on the underlying SUSY spectrum, in case a signal is
  seen.
\end{enumerate}

This last feature is particularly interesting for TGM models. From
the general discussion in appendix \ref{app:razor} we see that as far
as our spectrum is characterized by two well defined mass scales,
namely corresponding to 
$\tilde{q}, \tilde{u}^c$ and
$\tilde{d}^c$ squarks, 
the distribution of the $M_R$
variable will identify the latter as two different peaks of definite
mass. More specifically such peaks will occur for those values of
$M_\Delta$, see eq.~\eqref{eq:rzr_Mdelta}, corresponding to the
decays of the squarks towards the NLSP. The peculiar phenomenological
prediction of minimal unified TGM, the ratio in equation
\eqref{eq:sfermionmasses} would then be translated to a ratio between
the position of the two peaks in the distribution of $M_R$ given by
\begin{equation}
\frac{M_{\Delta}^{d^c, l}}{M_{\Delta}^{q, u^c, e^c}} = \sqrt{2} \left( 1 + \frac{m^2_{\rm{NLSP}}}{2 {m}^2_{10}}  + \dots \right) \, .
\end{equation}
Unfortunately the situation just depicted is too simplistic as many
different effects tend to broaden the $M_R$ distribution, causing a
partial or total overlap of the different peaks. Anyway, with high
luminosity and sufficient separation ($\gtrsim 30\%$ of the peak
position) one could distinguish the peaks even in presence of detector
effects.

\subsection{Analysis of the benchmark points}

We start by computing the SUSY spectrum evolving the parameters of
eqs.~\eqref{eq:W} and \eqref{eq:LSB} with the RGEs described in
appendix \ref{sec:RGE} down to low energies using a modified version
of the {\tt softSUSY} package \cite{Allanach:2001kg}; knowing the
spectrum we calculate the branching ratios via {\tt SUSY-HIT}
\cite{Djouadi:2006bz}.  Then we generate a sample of SUSY events at
the center of mass energy of 7 TeV using {\tt
  PYTHIA8}~\cite{pythia8}. We cluster jets from the stable particles
in the event, ignoring neutrinos and the NLSP, with the {\it anti-Kt}
jet algorithm~\cite{antiKt} as implemented in {\tt
  FASTJET}~\cite{fastjet1,fastjet2}.  The energy of the generator
level jets is then modified in order to take into account the detector
resolution of the CMS detector~\cite{PhysicsTDR}. The resolution is
modeled according to a Gaussian response function both for the jet
transverse momenta and the missing transverse energy (MET).

Our emulation of the CMS razor analysis follows the guideline provided
by the CMS collaboration~\cite{RazorCMSTwiki}.  We emulate the
performances of the CMS detector according to the provided
instructions before applying the analysis selection. We use the events
surviving the selection to build the 2D $R^2$ vs $M_R$ distributions
for the six exclusive {\it boxes}, which are used to derive a limit on
the cross section for a given SUSY model. The limit is computed
running the code provided by the CMS collaboration, which combines the
six exclusive boxes and incorporate the uncertainty on the
signal and the background distributions. It is interesting to
compare the distribution of $M_R$ and $R^2$ in different boxes. In
case of an observation, the prominence of the different
$M_R$ peaks in different boxes could be used to understand which
sparticles could have been produced in the collision or in the decay.

\bigskip

\begin{figure}
\begin{center}
\includegraphics[width=0.7\textwidth]{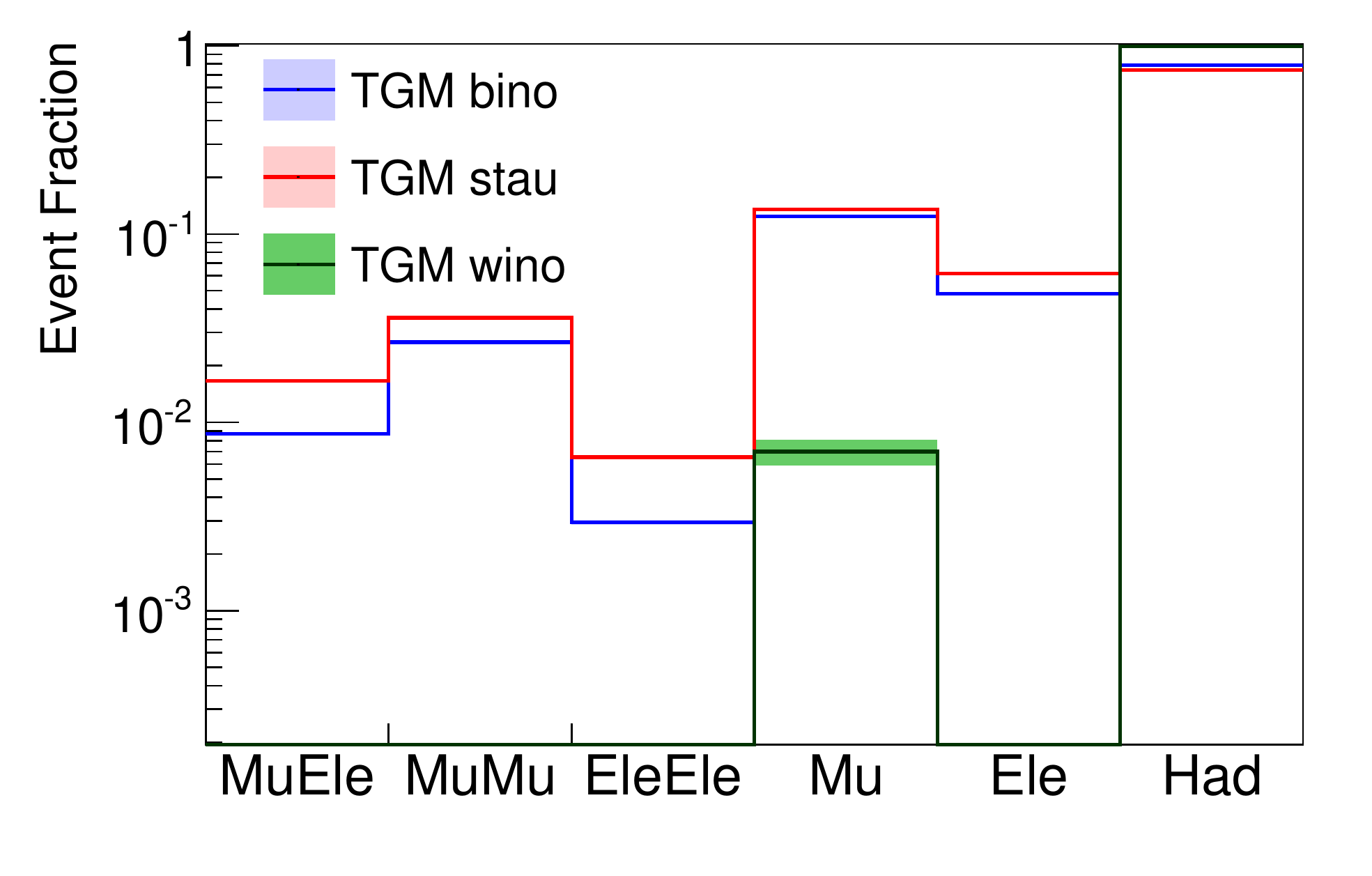}
\caption{\label{fig:boxBreakdown} Relative fraction of signal events
  in the six razor boxes, for the three considered benchmark models.}
\end{center}
\end{figure}

Different boxes are differently important for different models (see
Fig.~\ref{fig:boxBreakdown}). For instance, the wino benchmark model
is characterized by the production of ewkinos, which are too close in
mass for the model to be observable. In this case, SUSY production is
accessible only through the associated jet production, which explains
why the events fraction in the hadronic box is very close to one. For
the other models the event fraction in the hadronic box goes down to
$\sim 80\%$, while $\sim 5-10\%$ of the events fill the single-lepton
and the MuMu boxes. Given the larger background contamination in the
Had box, a larger yield does not necessarily correspond to a better
signal-to-background discrimination. 

\bigskip

\begin{figure}
\begin{center}
\includegraphics[width=0.48\textwidth]{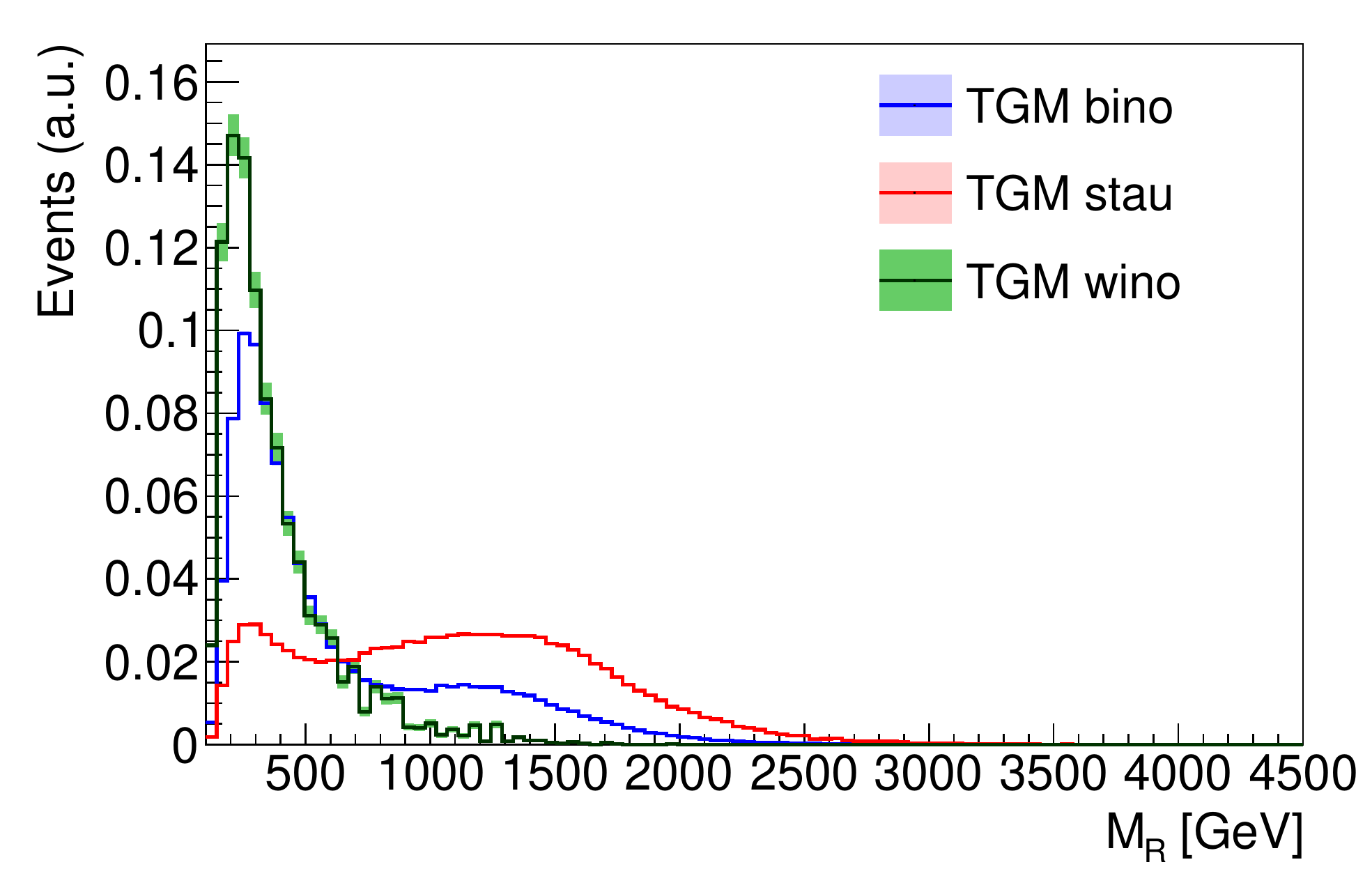}
\includegraphics[width=0.48\textwidth]{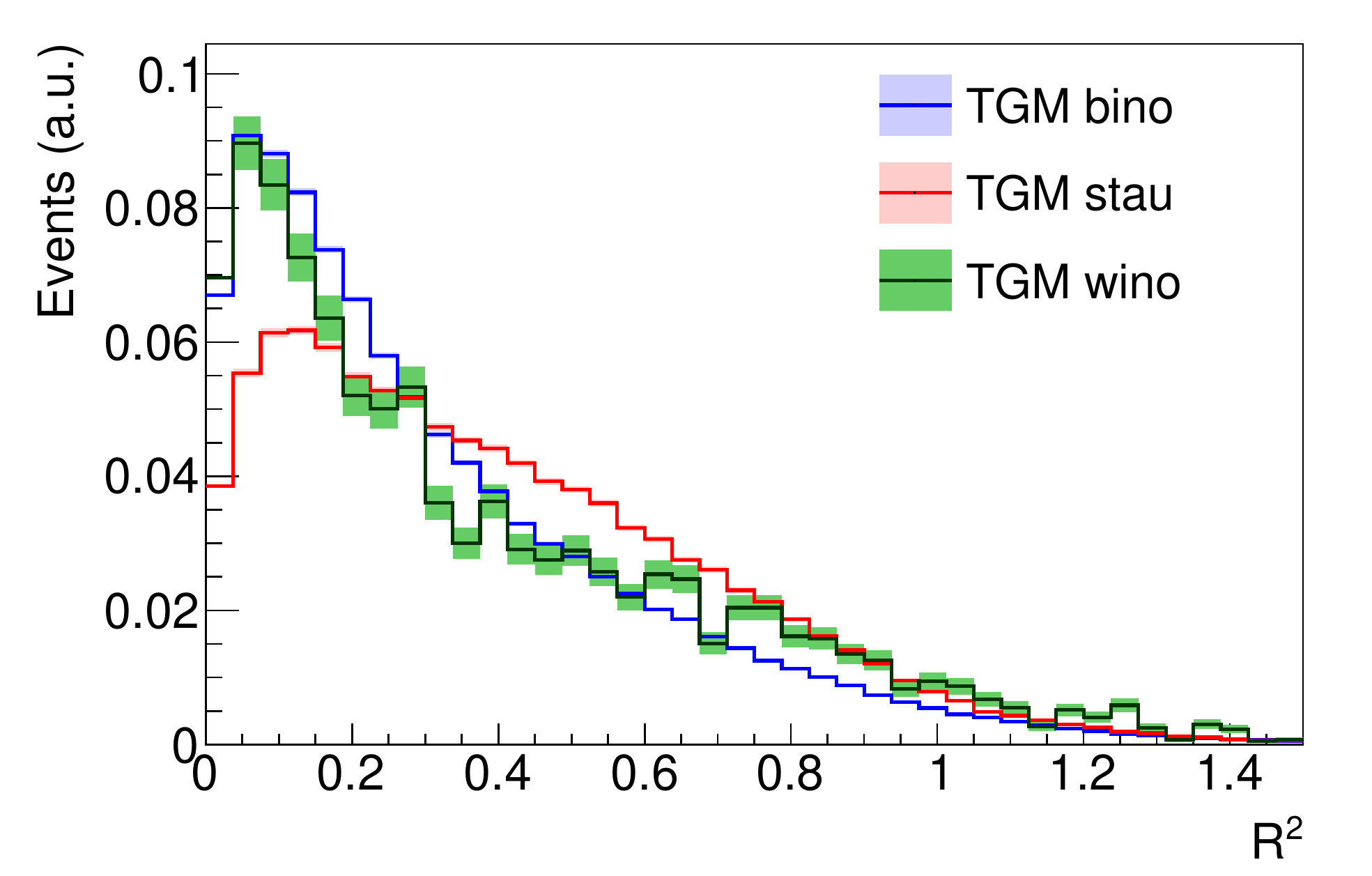} \\
\includegraphics[width=0.48\textwidth]{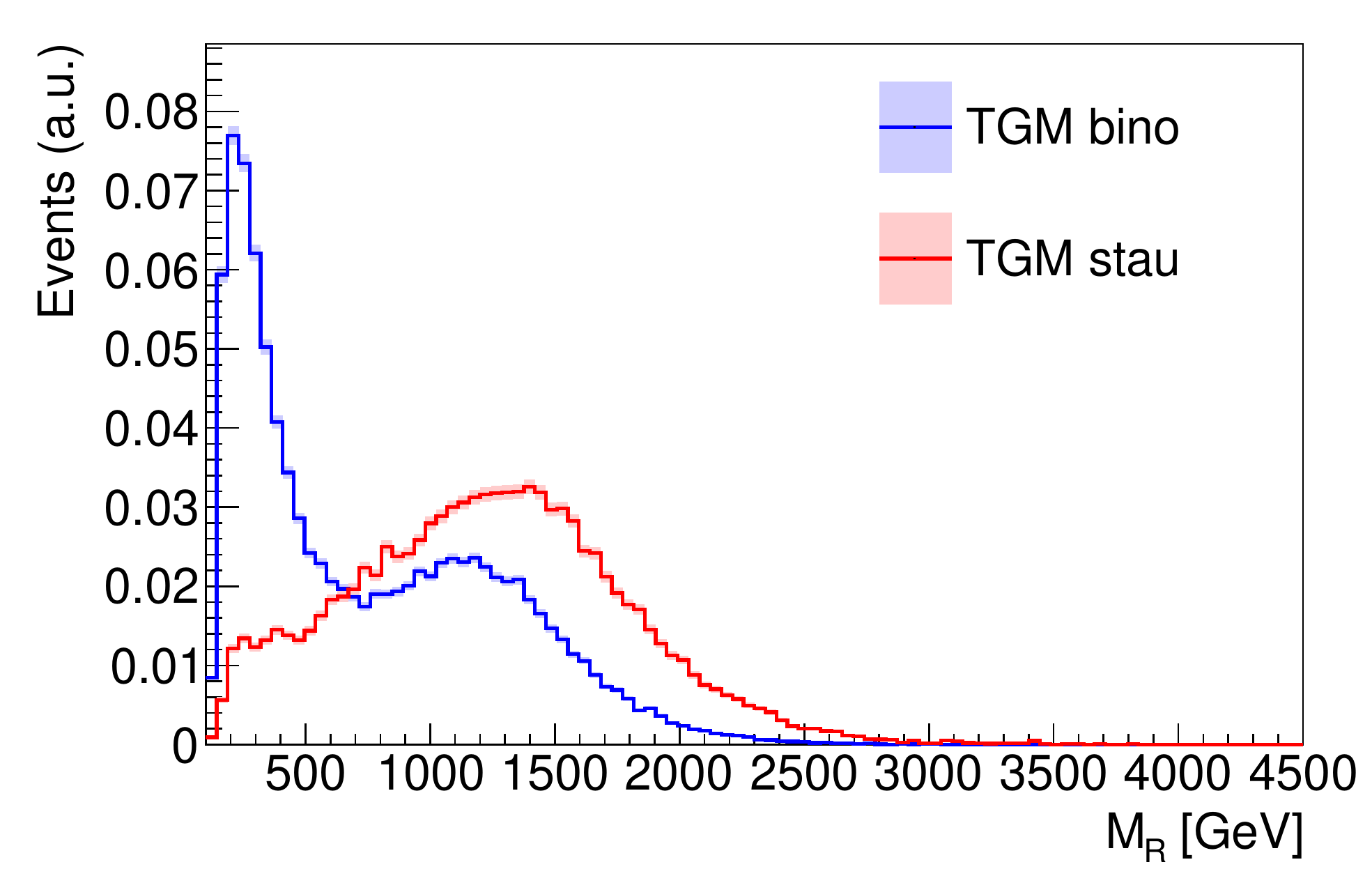}
\includegraphics[width=0.48\textwidth]{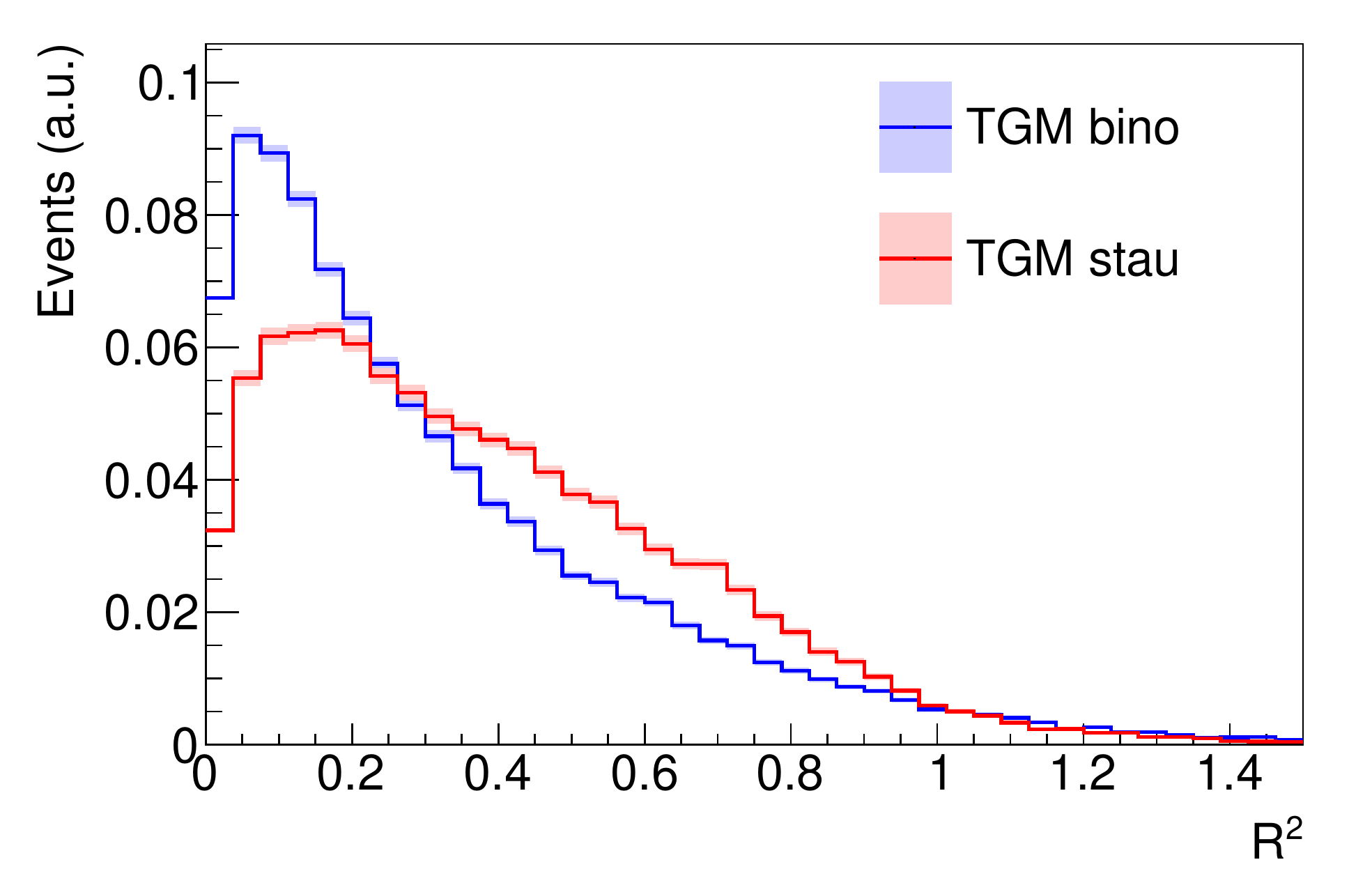} \\
\includegraphics[width=0.48\textwidth]{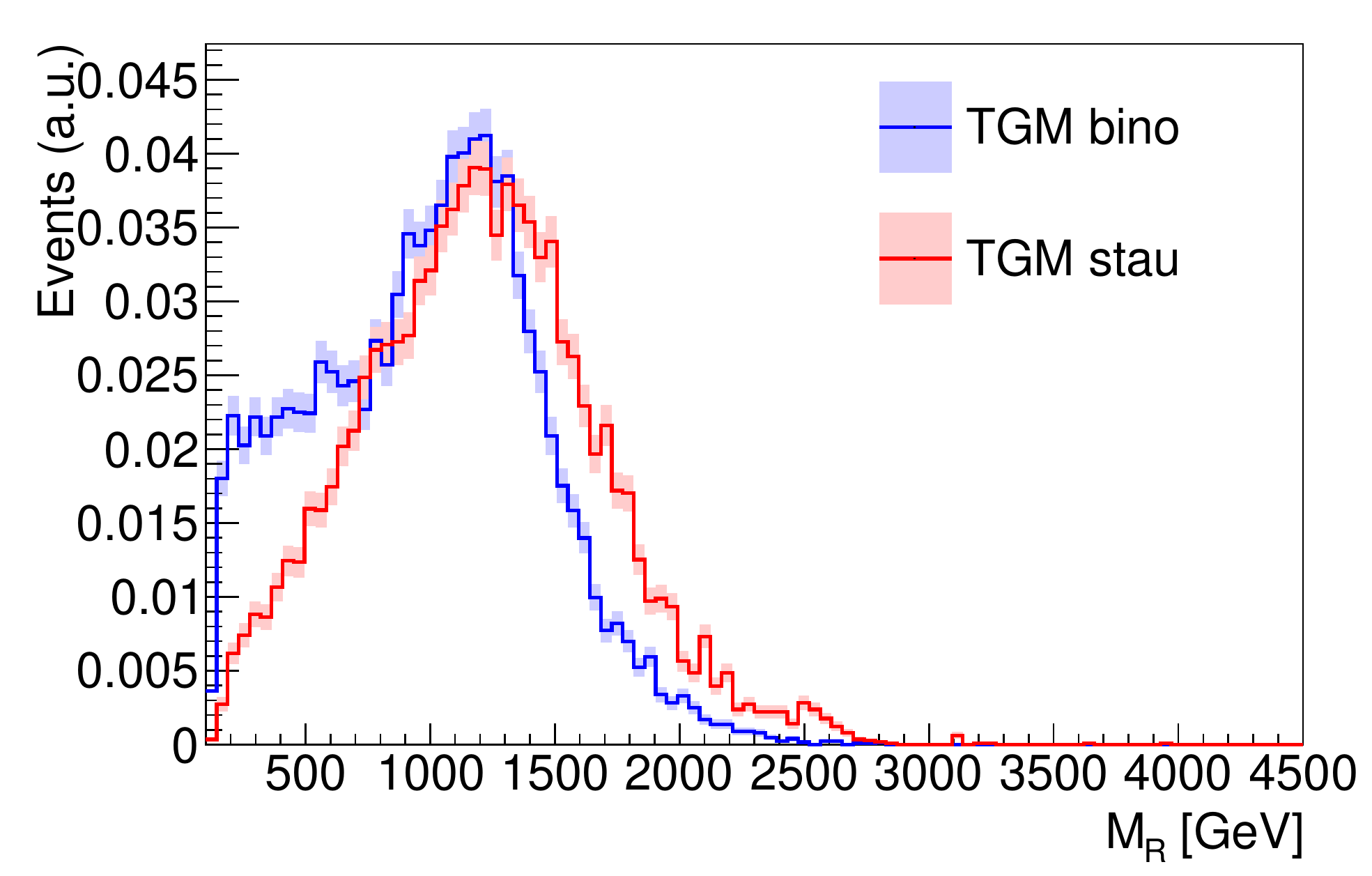}
\includegraphics[width=0.48\textwidth]{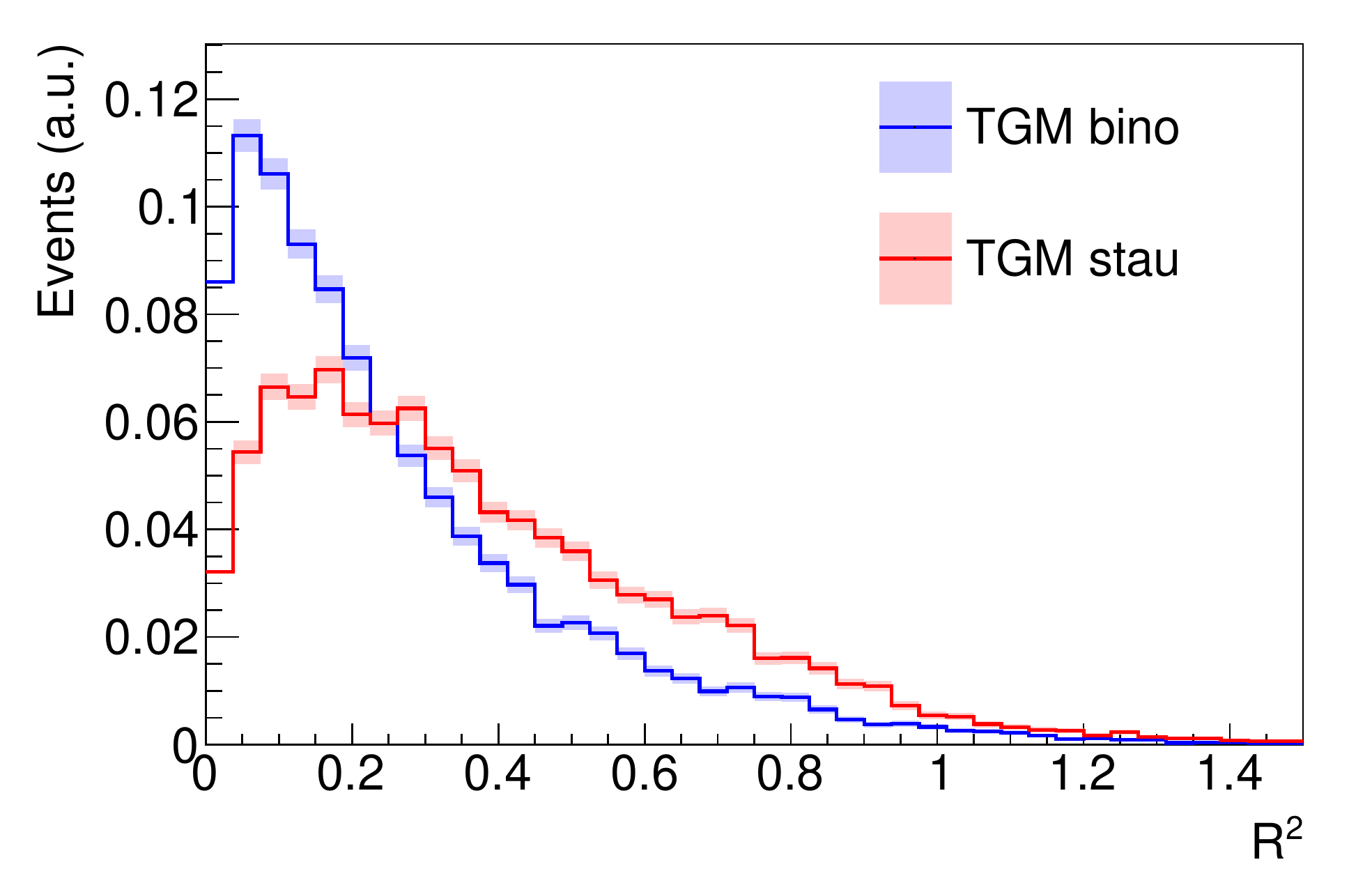} 
\caption{\label{fig:TGM_LHC_Boxes} $M_R$ (left) and $R^2$ (right)
  distributions for a set of TGM benchmark points, as obtained for the
  CMS razor hadronic (top), leptonic (center), and dileptonic (bottom)
  boxes.}
\end{center}
\end{figure}

We show in figure \ref{fig:TGM_LHC_Boxes} the $M_R$ and $R^2$
projections for the hadronic, leptonic, and semileptonic boxes in the
benchmark points under analysis. One could notice that the different
decay chains produce different distributions, even within one
model. The presence of two competitive decay chains in one model
generates a multimodal distribution, each local maximum corresponding
to a different mass split between the produced sparticle and the
NLSP. One should notice that we further assume the stable staus to be
too slow to be detected with the ordinary event
reconstruction\footnote{Recently, it was also pointed out that these
  particles could receive a boost if produced in the cascade decay of
  heavier particles. In this case they should be detected as ordinary
  muons, with no missing energy in the event. In this sense, any
  conclusion we obtain neglecting this effect overestimates the
  sensitivity of the razor analysis to these models, since a
  misidentification of the stau as a muon would reduce the value of
  $R^2$ and consequently the efficiency of the analysis.}.

The $M_R$ distribution is characterized by two peaks. The broad peak
around 1 TeV in the hadronic box for the stau and the bino models is
the overlap of the competing gluino-gluino and squark-gluino
production mechanisms. Due to the resolution in $M_R$ and the small
mass differences between the squarks and the gluino, it is not
possible to resolve the different peaks. This peak is also present in
the leptonic boxes, the lepton being produced in the cascade decays of
the squarks, typically from $W$ and $Z$ bosons coming from
ewkinos. 

The second peak at low $M_R$ has a different origin. The events
around this peak originate from the production of charginos and
neutralinos. Being very close in mass, these particles tend to produce
soft objects (jets or leptons) when decaying to the NLSP. These
events are in general rejected by the event selection, which requires
two jets with a transverse momentum of at least 60~GeV, unless the
charginos and neutralinos are produced in association with at least
two jets coming from initial or final state radiation. In this case, the
visible jet and the invisible massive particles do not originate from the
decay of a heavy sparticle, as the razor construction assumes. These
events correspond to a non resonant production and no peak in $M_R$
is expected. If the jet $p_T$ requirement was lower, one would see a
falling distribution for $M_R$. On the other hand, only events with two
energetic jets enter the distribution. These events have an intrinsic
requirement on the minimum visible energy of the event, which (due to
the correlation between $M_R$ and the visible energy) scalps the
$M_R$ distribution at low values, producing what looks like a peak at
low $M_R$. Unlike the case of genuine kinematic peaks, the position
of this peak is not related to the SUSY spectrum, being a
model-independent artifact of the event selection.
The abundance of these
events is maximal for hadronic events and reduced for one-lepton
events, while it become subdominant for two-lepton events. In the case
of the wino benchmark point the split in mass between the chargino and
the neutralino is so small that the leptons are undetected in the
majority of the cases. As a consequence, almost all the events fall in
the hadronic box. The relative importance of the two contributions in
different boxes could give an insight of the relative cross sections
for the two classes of process, which eventually could allow to
constrain the mass scale associated to the produced particles.

Following the instructions given by CMS~\cite{RazorCMSTwiki} we
compute the excluded cross section for each benchmark model and
compare it to the next to leading order (NLO) value, obtained running
{\tt PROSPINO}~\cite{PROSPINO}. In the case of the stau benchmark
point one would need a more detailed detector simulation to correctly
take into account the fraction of events in which the two staus
actually contribute to the missing transverse energy in the event. If
this fraction is small, the limit would be much weaker than what is
quoted in Table \ref{chap:TGM_LHC_tabXsec}.

\begin{table}
\begin{center}
\begin{tabular}{cccc}
\toprule
Model & NLO SUSY  & Had-box excluded  & Total excluded \\
 & cross section [pb] & cross section [pb] & cross section [pb] \\
\midrule
TGM bino & 0.027 & 0.024 & 0.019\\
TGM wino & 12.02 & 4.3 & 3.5 \\
TGM stau & 0.002 & 0.010 & 0.008 \\
\bottomrule
\end{tabular}
\caption{Theoretical NLO SUSY cross section for the three benchmark
  points obtained from {\tt PROSPINO}~\cite{PROSPINO} compared to the
  excluded cross section (at 95\% probability) estimated with our
  implementation of the razor analysis by the CMS collaboration,
  according to the procedure given by the CMS collaboration
  \cite{RazorCMSTwiki}. Both the limit from the Hadronic box and the
  combined limit are shown. }
\label{chap:TGM_LHC_tabXsec}
\end{center}
\end{table}

The largest sensitivity comes from the hadronic box, which collects
the majority of the events originating from the production of colored
sparticles. The improvement due to the leptonic boxes is marginal for
the considered benchmark models.  The stau and the bino models are not
excluded. But the observed limit is not far from the model cross
section, such that the analysis of the 8 TeV data could already rule
them out. The wino point is excluded, despite being the most
challenging. This proves that the cross section production for ewkinos
lighter than 200 GeV is already probed by the 7 TeV LHC data, the
cross section being above 1 pb.
Additional sensitivity could be provided by dedicated searches for
directly-produced charginos and neutralinos. The exclusion reach by
the ATLAS~\cite{ATLASEW} and CMS~\cite{CMSEW} multilepton analyses,
obtained considering the full 8 TeV statistics, is not good enough to
cover the benchmark models we considered. This is mainly due to the
large chargino and neutralino masses and the corresponding suppression
of the production cross section. These benchmark models could be probed
with the next LHC run, thanks to the larger production cross section
and the larger expected statistics.

\bigskip

\begin{figure}
\centering
\includegraphics[scale=0.8]{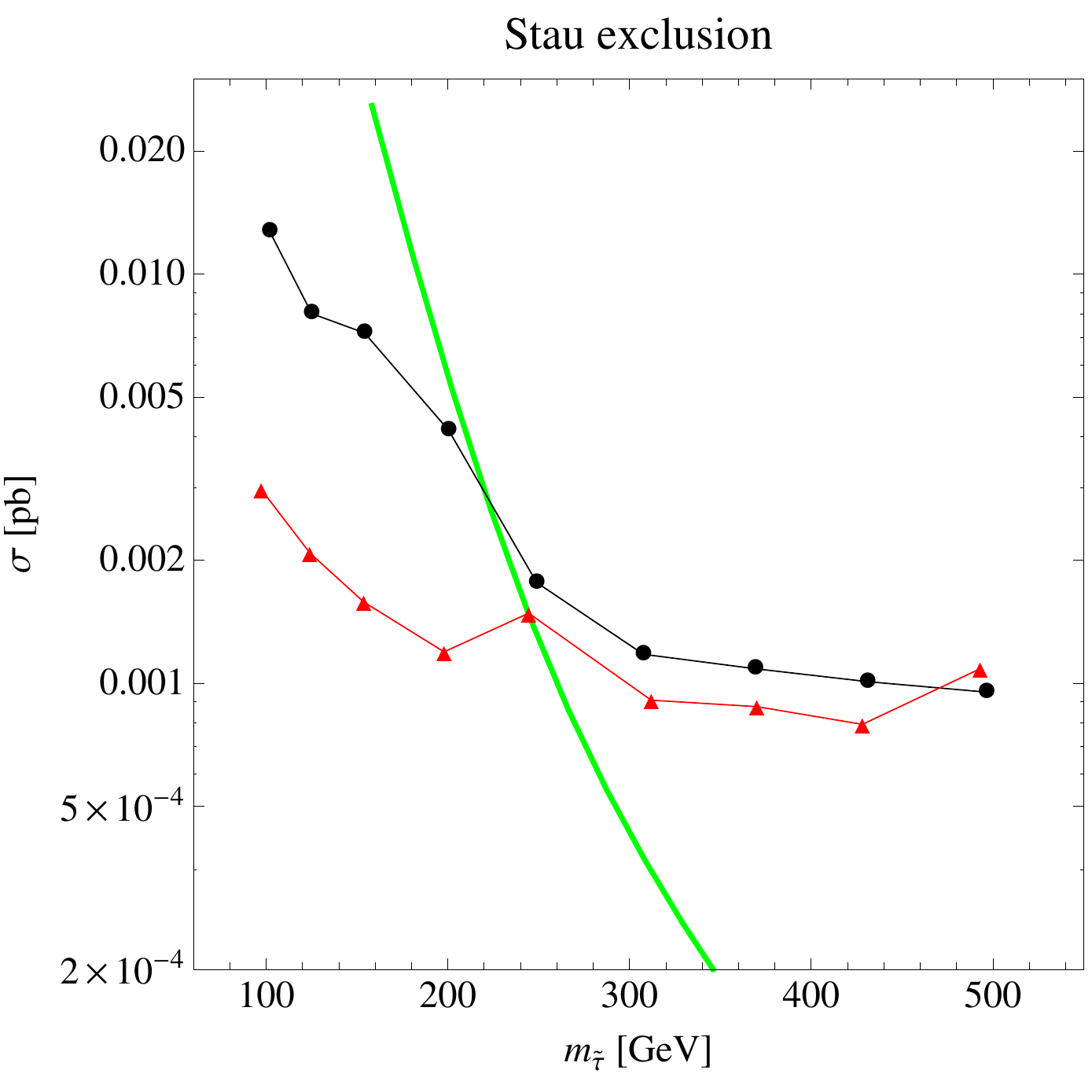} 
\caption{In order to set some limits on the NLSP stau mass we calculated the predicted theoretical cross section and then compared the latter with the observed 95\% CL upper limit \cite{heavychargedexp}. The black line represents the experimental bound on the cross section taking  into account only the selection based on the tracker, while the red line is based also on the time of flight (TOF). The green line gives the theoretical direct production cross sections for staus on which we added the subleading contribution of the indirect stau production owed to the squark and gluino channels, all of these contribution computed through {\tt PROSPINO} \cite{PROSPINO}. All in all we can give a mass bound for the stau of 220 $\div$ 250~GeV.}
\label{fig:staulim}
\end{figure}

Finally, in the case in which the NLSP is the stau some bounds on its mass can be set from the searches on the heavy charged stable particles, as anticipated in section \ref{chap:TGM_LHC_benchs_stau}: the stau, indeed, decays to the gravitino outside the detector. Such limits in the TGM framework are in general less restrictive than those in MGM since the additive tree level contribution to stau soft mass term accounts for a comparably smaller production cross section.
As shown in figure \ref{fig:staulim}, the recent experimental results allows a TGM stau mass larger than 220 $\div$ 250~GeV. In that plot, we have varied $m_{10}$ from 450 to 1250~GeV and fixed the other parameters at the values of the stau benchmark point.

\section{Summary and conclusions}

We studied the LHC phenomenology of a minimal unified realization of Tree-level Gauge Mediation, in particular the possibility to test its peculiar prediction for the sfermion mass ratios. We did this in three steps. 

First, we provided a detailed definition of the Lagrangian of the model and of the relevant parameters, taking into account the possible deviations from $SO(10)$ relations due to the contributions of the non-renormalizable operators necessary to fix the GUT prediction for the light fermion mass ratios. 

Then we discussed the precise determination of the low energy spectrum in terms of the above parameters. In particular, we provided analytical formulas for the RGE running and a numerical implementation in {\tt softSUSY}. The possible deviations from  GUT relations turn out not to affect the tree-level predictions for the sfermion mass ratios. However, we pointed out that they can give rise to largely non-universal gaugino masses without any conflict with the unification of gauge couplings. The non-universality arises from the flavour structure of the messenger interactions. Even in the limit in which the $SO(10)$ breaking effects are small and only significantly affect the small Yukawa couplings of the first families, the effect on gaugino masses can be sizeable. This is because gaugino masses are equally sensitive to the ratio of the larger third family Yukawa couplings and to the ratio of the smaller first family couplings, more likely to be affected by $\ord{1}$ effects. As a consequence of the possible non-universality of gaugino masses, the Wino can be lighter than the Bino. Still, gaugino masses satisfy a sum rule, \eq{sumrule}, which can be considered as another smoking gun of minimal unified TGM. 

Another important aspect related to the determination of the TeV-scale spectrum has to do with $A$-terms. Usually in gauge mediation no $A$-terms are generated at the one-loop level at the messenger scale. This is not the case here. In fact, the MSSM fields and the minimal gauge mediation messengers lie in the same $SO(10)$ multiplets, so that the messenger-messenger-$Z$ coupling generating gaugino masses are accompanied by matter-messenger-$Z$ couplings generating non-vanishing $A$-terms at the messenger scale. The size of the $A$-terms depends on whether the heavier of the three families of messengers is significantly lighter than the GUT scale or not. The latter case, corresponding to third family Yukawas of order one in the full $SO(10)$ theory, gives rise to larger  $A$-terms but is more model-dependent, as it depends on unknown details of the full $SO(10)$ theory. In turn, the possibility of sizeable $A$-terms allows to account for a $125\GeV$ light Higgs for sfermion masses within the LHC reach. On the other hand, the Higgs mass can be raised above the MSSM prediction with a suitable implementation of the NMSSM setup. Another interesting property of the $A$-terms in TGM is that only the third family $A$-terms are non negligible, as the first and second family are suppressed by powers of small Yukawa couplings. This solves the supersymmetric CP problem. 

Different possible types of spectra can be obtained, in particular as far as the NLSP is concerned. In TGM models, the Lightest Supersymmetric Particle is the gravitino. The NLSP turns out to be a bino-like neutralino, a wino-like neutralino, or a stau. The possibility that the lightest neutralino is wino like is opened by the possible non-universality of gaugino masses. We have therefore considered three benchmark points representative of each of those possibilities. 

Finally, we studied the LHC signals associated to each benchmark point, considering in particular the possibility to test the sfermion mass ratio prediction. TGM turns out to be an interesting playground on which the performances of different searches can be compared. From this point of view we found that the razor inclusive analysis by CMS was an ideal tool. 

For each benchmark point, we computed the SUSY spectrum by running the
parameters of the model from the GUT to the TeV scale using a modified
version of the {\tt softSUSY} package. We evaluated the possible
signatures at the LHC applying the selection of the CMS Razor
analysis, discussing the interesting interplay between the different
production processes and decay chains in the different {\it boxes} in
which the Razor search is defined. We also studied other
interesting features of the TGM benchmark models, as for instance the
long-living staus, the compressed chargino-neutralino spectrum and the
large mass difference between the colored particles and the rest of
the spectrum. The TGM class of models can accomodate the lack of a
SUSY signal so far and the possibility of observing one with the 8 TeV
data, or with the first data collected at higher energy at the LHC
restart.

\section*{Acknowledgments}
The work of A.R.\ and M.S.\ was supported by the ERC Advanced Grant no. 267985 ``DaMESyFla'', by the EU Marie Curie ITN ``UNILHC'' (PITN-GA-2009-237920) and the European Union FP7 ITN invisibles (Marie Curie Actions, PITN- GA-201-289442). Part of this work was done at the Galileo Galilei Institute for Theoretical Physics, which we thank for the kind hospitality and support.

\appendix

\section{Flavour structure of the superpotential}
\label{sec:parameters}

In this appendix, we discuss the expectations for the size of the parameters $c_{D_i,L_i}$. As we have seen in section \ref{sec:messmass}, the breaking of $SO(10)$ and SUSY must involve spinorial representations. In particular, the $\mathbf{16}$, $\overline{\mathbf{16}}$ fields acquire a vev $M$ in the scalar, SM singlet component and $\mathbf{16}'$, $\overline{\mathbf{16}}'$ acquire a vev in the $F$-term SM singlet component. As in section \ref{sec:messmass}, we will actually assume for simplicity that only $\mathbf{16}'$ gets an $F$-term and we further assume that the $\mathbf{16}$ and $\mathbf{16}'$  are the only spinorial representations coupling to matter bilinears. For convenience we remind the form of the most general R-parity invariant superpotential bilinear in the matter fields $\mathbf{16}_i+\mathbf{10}_i$, \eq{W2}:
\begin{equation}
\label{eq:W2app}
W_2 = h_{ij} \mathbf{16}_i \mathbf{10}_j \mathbf{16} + h'_{ij} \mathbf{16}_i \mathbf{10}_j \mathbf{16}' + \frac{y_{ij}}{2} \mathbf{16}_i \mathbf{16}_j \mathbf{10} + W_2^\text{NR}.
\end{equation}
In the previous equation, a mass term $\mu_{ij} \mathbf{10}_i\mathbf{10}_j$ has been assumed to be absent to obtain a ``pure'' embedding of the SM fields in $SO(10)$ representations and to avoid reintroducing the flavour problem~\cite{TGM2}. The (model-dependent) non-renormalizable part is not specified but it is supposed to bring the fermion mass ratios to the phenomenologically correct values.

In order to identify the light Yukawa couplings we need to specify better the embedding of the light Higgs fields, deepening the discussion in section \ref{sec:RelevantParameters}. The light $h_d$ can  be contained in the doublet component of the $\mathbf{16}$, $h^{16}_d$, in the doublet of the $\mathbf{16}'$, $h^{16'}_d$ or in a $\mathbf{10}$, with the size of the total component in spinorial representations given by $\sin\theta_d$. The field $h_d$ could be in principle also be embedded in a spinorial representation different from $\mathbf{16}$ and $\mathbf{16}'$ and not coupling to the matter bilinears, but we assume that this is not the case. We can use an angle $\alpha$ to measure how $h_d$ is shared by the two spinorial representations:
\begin{equation}
\label{eq:KK}
h^{16}_d = \sin\theta_d \cos\alpha\, h_d +\text{\ldots} \,, \quad
h^{16'}_d = \sin\theta_d \sin\alpha\, h_d +\text{\ldots}\,.
\end{equation}
From eqs.~\eqref{eq:W2app} and \eqref{eq:KK} we can recover the SM Yukawa couplings $\lambda_{U,D,E}$ and $\hat{\lambda}_{D,E}$ in eq.~\eqref{eq:W} as follows: 
\begin{equation}
\label{eq:chap_TGMeq:yukawas}
 \begin{aligned}
 \lambda_U &= \cos\theta_u\, y + \lambda^\text{NR}_U \;, & & \\
\lambda_E &= \sin \theta_d (\cos \alpha h + \sin \alpha h') + \lambda_E^\text{NR} \;,
&
\lambda_D &= \sin \theta_d (\cos \alpha h + \sin \alpha h') + \lambda_D^\text{NR} \;, \\
\hat{\lambda}_E &= \cos \theta_d\,y + \hat{\lambda}_E^\text{NR} \;,
&
\hat{\lambda}_D &= \cos \theta_d\,y + \hat{\lambda}_D^\text{NR} \;, \\
\end{aligned}
\end{equation}
where the superscript "NR" denotes a correction vanishing in the limit $W^\text{NR}_2 \to 0$.

From \eqs{chap_TGMeq:yukawas} one can see that the simplest possible relation between the  parameters $h_{D,L}$ and the MSSM Yukawas is obtained when $h_d$ is entirely in the $\mathbf{16}$ and the non-renormalizable contributions are negligible, in which case we obtain \eq{minimalpars}. In order to account for the general case, we have introduced new parameters $c_{D_i,L_i}$ defined by
\begin{equation}
\label{eq:cpars2}
\begin{aligned}
{h_D}_i &= {c_D}_i {\lambda_D}_i / \sin\theta_d  \;,
&\qquad \qquad
{h_L}_i &= {c_L}_i {\lambda_L}_i / \sin\theta_d  \;, 
\end{aligned}
\end{equation}
The $c_{D_i,L_i}$ coefficients can be written in terms of the parameters in eqs.~\eq{chap_TGMeq:yukawas} as follows:
\begin{equation}
\label{eq:cdl}
\begin{aligned}
c_{L_i} = \frac{1}{\cos \alpha + \sin \alpha \,{\gamma_L}_i} + (c_{L_i})_\text{NR}, \qquad
c_{D_i} = \frac{1}{\cos \alpha + \sin \alpha\, {\gamma_D}_i} + (c_{D_i})_\text{NR} .
\end{aligned}
\end{equation}
The equations above allow to set an appropriate range for these coefficients. In the limit in which $h_d$ lies in the $\mathbf{16}$ only ($\alpha = 0$), $c_{D_i,L_i} = 1$ at the renormalizable level. In the limit in which $h_d$ lies in the $\mathbf{16'}$ only ($\alpha = \pi/2$), on the other hand, the parameters $c_{D_i,L_i}$ can be smaller, especially if the parameters $\gamma_{D,L}$ in \eqref{eq:gammas} enhance gaugino masses. 

\section{One-loop RGEs}
\label{sec:RGE}

In this section we shall present the RGEs for the full theory below the GUT scale \cite{RGEsusy}. In all of the following equations we will use the common definition $t \equiv \ln \mu$ where $\mu$ is the renormalization scale.

\subsection{Gauge couplings}

The RGEs for the gauge couplings are
\begin{equation}
(4 \pi)^2 \frac{\text{d} g_a}{\text{d} t} = \beta^{(1)}{g_a} \,,
\end{equation}
where
\begin{equation}
\beta^{(1)}{g_a} = 
g_a^3 \sum_R B_a(R)
\end{equation}
and 
\globallabel{1}
\begin{align}
B_3 &= \sum_R B_3(R) = - 3 + \frac{N_{D^c} + N_{D}}{2} \,,\mytag\\
B_2 &= \sum_R B_2(R) = 1 + \frac{N_{L} + N_{L^c}}{2} \,,\mytag\\
B_1 &= \sum_R B_1(R) = \frac{33}{5} + \frac{3}{5} \Bigl(\frac{1}{3} N_{D^c} + \frac{1}{3} N_{D} + \frac{1}{2} N_{L} + \frac{1}{2} N_{L^c} \Bigr) \,, \mytag
\end{align}
where $N_{D^c}$ is the number of $D^c$ fields and similar for the other $N$.

\subsection{Gaugino masses}

In terms of the results obtained for the gauge couplings one has
\begin{equation}
(4 \pi)^2 \frac{\text{d} M_a}{\text{d} t} = 2 g_a^2 B_a M_a \,.
\end{equation}

\subsection{Yukawa couplings}

In the following equations, the integration of the heavy chiral messengers at their mass scale is taken into account by setting to zero the corresponding entries of the Yukawa matrices.  We note that the part proportional to the gauge coupling does not depend on the number of flavours that are switched on since it is directly related to the specific $\lambda$ parameter under study. Incidentally we note that if some of the flavours are frozen out this will also act on the meaning of the various traces appearing in the equations.
\globallabel{2}
\begin{align}
(4 \pi)^2 \frac{\text{d} \lambda_U}{\text{d} t} &= \lambda_U \Bigl[ \tr (3 \lambda_U^{\dagger} \lambda_U) + 3 \lambda_U^{\dagger} \lambda_U + \lambda_D^{\dagger} \lambda_D + {\hat{\lambda}_D}^{\dagger} \hat{\lambda}_D - \frac{16}{3} g_3^2 - 3 g_2^2 - \frac{13}{15} g_1^2 \Bigr] \mytag\\
(4 \pi)^2 \frac{\text{d} \lambda_D}{\text{d} t} &= \lambda_D \Bigl[ \tr (3 \lambda_D^{\dagger} \lambda_D + 3 {\hat{\lambda}_D}^{\dagger} \hat{\lambda}_D + \lambda_E^{\dagger} \lambda_E + {\hat{\lambda}_E}^{\dagger} \hat{\lambda}_E) + 3 \lambda_D^{\dagger} \lambda_D + 3 {\hat{\lambda}_D}^{\dagger} \hat{\lambda}_D +  \lambda_U^{\dagger} \lambda_U \Bigr] \nonumber \\ 
& {} \quad {} - \lambda_D \Bigl[ \frac{16}{3} g_3^2 + 3 g_2^2 + \frac{7}{15} g_1^2 \Bigr] \mytag\\
(4 \pi^2) \frac{\text{d} \lambda_E}{\text{d}t} &= \Bigl[ \tr (3 \lambda_D^{\dagger} \lambda_D + 3 {\hat{\lambda}_D}^{\dagger} \hat{\lambda}_D + \lambda_E^{\dagger} \lambda_E + {\hat{\lambda}_E}^{\dagger} \hat{\lambda}_E) + 3 \lambda_E \lambda_E^{\dagger} + 3 {\hat{\lambda}_E} \hat{\lambda}_E^{\dagger} \Bigr] \lambda_E \nonumber \\
& {} \quad {}- \Bigl[ 3 g_2^2 + \frac{9}{5} g_1^2 \Bigr] \lambda_E \mytag\\ 
(4 \pi^2) \frac{\text{d}\hat{\lambda}_D}{\text{d}t} &= \hat{\lambda}_D \Bigl[ \tr (3 \lambda_D^{\dagger} \lambda_D + 3 {\hat{\lambda}_D}^{\dagger} \hat{\lambda}_D + \lambda_E^{\dagger} \lambda_E + {\hat{\lambda}_E}^{\dagger} \hat{\lambda}_E) + 3 {\hat{\lambda}_D}^{\dagger} \hat{\lambda}_D + 3 \lambda_D^{\dagger} \lambda_D +  \lambda_U^{\dagger} \lambda_U \Bigr] \nonumber \\
& {} \quad {}- \hat{\lambda}_D \Bigl[ \frac{16}{3} g_3^2 + 3 g_2^2 + \frac{7}{15} g_1^2 \Bigr] \mytag\\
(4 \pi^2) \frac{\text{d}\hat{\lambda}_E}{\text{d} t} &= \Bigl[ \tr (3 \lambda_D^{\dagger} \lambda_D + 3 {\hat{\lambda}_D}^{\dagger} \hat{\lambda}_D + \lambda_E^{\dagger} \lambda_E + {\hat{\lambda}_E}^{\dagger} \hat{\lambda}_E) + 3 \lambda_E \lambda_E^{\dagger} + 3 {\hat{\lambda}_E} \hat{\lambda}_E^{\dagger} \nonumber\\
& {} \quad {} - 3 g_2^2 - \frac{9}{5} g_1^2 \Bigr] \hat{\lambda}_E \mytag
\end{align}

\subsection[The $\mu$ parameter and other bilinear terms in the superpotential]{The $\boldsymbol{\mu}$ parameter and other bilinear terms in the superpotential}
\label{app:TGM_muparameter}

The running of the dimension one parameters in the superpotential is given by
\globallabel{3}
\begin{align}
(4 \pi)^2 \frac{\text{d} \mu}{\text{d} t} &= \mu \Bigl[ \tr (3 \lambda_D^{\dagger} \lambda_D + 3 {\hat{\lambda}_D}^{\dagger} \hat{\lambda}_D + 3 \lambda_U^{\dagger} \lambda_U + \lambda_E^{\dagger} \lambda_E + {\hat{\lambda}_E}^{\dagger} \hat{\lambda}_E) - 3 g_2^2 - \frac{3}{5} g_1^2 \Bigr]  \mytag\\
(4 \pi)^2 \frac{\text{d} M_{D}}{\text{d} t} &= 2 \hat{\lambda}_D \Bigl( {\hat{\lambda}_D}^{\dagger} M_{D} + {\lambda_D}^{\dagger} M_{d D} \Bigr) - \Bigl( \frac{16}{3} g_3^2 + \frac{4}{15} g_1^2 \Bigr) M_{D}   
\mytag \\%
(4 \pi)^2 \frac{\text{d} M_{d D} }{\text{d} t} &= 2 \lambda_D \Bigl( {\hat{\lambda}_D}^{\dagger} M_{D} + {\lambda_D}^{\dagger} M_{d D} \Bigr) - \Bigl( \frac{16}{3} g_3^2 + \frac{4}{15} g_1^2 \Bigr) M_{d D} 
\mytag \\%
(4 \pi)^2 \frac{\text{d} M_L}{\text{d} t} &= \hat{\lambda}_E^T \Bigl( \lambda_E^* M_{lL} + \hat{\lambda}_E^* M_L \Bigr) - \Bigl( 3 g_2^2 + \frac{3}{5} g_1^2 \Bigr) M_L 
\mytag \\%
(4 \pi)^2 \frac{\text{d} M_{l L}}{\text{d} t} &= \lambda_E^T \Bigl( \lambda_E^* M_{lL} + \hat{\lambda}_E^* M_L \Bigr) - \Bigl( 3 g_2^2 + \frac{3}{5} g_1^2 \Bigr) M_{l L}
\mytag 
\end{align}

\subsection{Trilinear SUSY breaking interactions}

Now we turn to the study of the SUSY breaking interaction terms of the Lagrangian. The running of the $A$-terms is given by 
\globallabel{4}
\begin{align}
(4 \pi)^2 \frac{\text{d} A_U}{\text{d} t} &= A_U \Bigl[ \tr (3 \lambda_U^{\dagger} \lambda_U) + 5 \lambda_U^{\dagger} \lambda_U + \lambda_D^{\dagger} \lambda_D + {\hat{\lambda}_D}^{\dagger} \hat{\lambda}_D - \frac{16}{3} g_3^2 - 3 g_2^2 - \frac{13}{15} g_1^2 \Bigl] \nonumber \\
&{}\quad{} + 2 \lambda_U \Bigl[ \tr (3 \lambda_U^{\dagger} A_U) + 2 \lambda_U^{\dagger} A_U + \lambda_D^{\dagger} A_D + {\hat{\lambda}_D}^{\dagger} \hat{A}_D \nonumber \\
&{}\quad{} + \frac{16}{3} M_3 g_3^2 + 3 M_2 g_2^2 + \frac{13}{15} M_1 g_1^2 \Bigl]   
\mytag \\%
(4 \pi)^2 \frac{\text{d} A_D}{\text{d} t} &= A_D \Bigl[ \tr (3 \lambda_D^{\dagger} \lambda_D + 3 {\hat{\lambda}_D}^{\dagger} \hat{\lambda}_D + \lambda_E^{\dagger} \lambda_E +  {\hat{\lambda}_E}^{\dagger} \hat{\lambda}_E) + 5 \lambda_D^{\dagger} \lambda_D + 5 {\hat{\lambda}_D}^{\dagger} \hat{\lambda}_D \nonumber \\ 
&{}\quad{} + \lambda_U^{\dagger} \lambda_U - \frac{16}{3} g_3^2 - 3 g_2^2 - \frac{7}{15} g_1^2 \Bigl]\nonumber \\
&{}\quad{} + 2 \lambda_D \Bigl[ \tr (3 \lambda_D^{\dagger} A_D + 3 {\hat{\lambda}_D}^{\dagger} \hat{A}_D + \lambda_E^{\dagger} A_E + {\hat{\lambda}_E}^{\dagger} \hat{A}_E) + 2 \lambda_D^{\dagger} A_D + 2 {\hat{\lambda}_D}^{\dagger} \hat{A}_D \nonumber \\  
&{}\quad{} + \lambda_U^{\dagger} A_U + \frac{16}{3} M_3 g_3^2 + 3 M_2 g_2^2 + \frac{7}{15} M_1 g_1^2 \Bigr]   
\mytag \\%
(4 \pi)^2 \frac{\text{d} A_E}{\text{d} t} &= A_E \Bigl[ \tr (3 \lambda_D^{\dagger} \lambda_D + 3 {\hat{\lambda}_D}^{\dagger} \hat{\lambda}_D + \lambda_E^{\dagger} \lambda_E +  {\hat{\lambda}_E}^{\dagger} \hat{\lambda}_E) + 5 \lambda_E^{\dagger} \lambda_E - 3 g_2^2 - \frac{9}{5} g_1^2 \Bigl] \nonumber \\ 
&{}\quad{} + 2 \lambda_E \Bigl[ \tr (3 \lambda_D^{\dagger} A_D + 3 {\hat{\lambda}_D}^{\dagger} \hat{A}_D + \lambda_E^{\dagger} A_E + {\hat{\lambda}_E}^{\dagger} \hat{A}_E) + 2 \lambda_E^{\dagger} A_E \nonumber \\
&{}\quad{}  + 3 M_2 g_2^2 + \frac{9}{5} M_1 g_1^2 \Bigr]  + 5 \hat{A}_E {\hat{\lambda}_E}^{\dagger} \lambda_E + 4 \hat{\lambda}_E {\hat{\lambda}_E}^{\dagger} A_E  
\mytag \\%
(4 \pi)^2 \frac{\text{d} \hat{A}_D}{\text{d} t} &= \hat{A}_D \Bigl[ \tr (3 \lambda_D^{\dagger} \lambda_D + 3 {\hat{\lambda}_D}^{\dagger} \hat{\lambda}_D + \lambda_E^{\dagger} \lambda_E +  {\hat{\lambda}_E}^{\dagger} \hat{\lambda}_E) + 5 \lambda_D^{\dagger} \lambda_D + 5 {\hat{\lambda}_D}^{\dagger} \hat{\lambda}_D \nonumber \\
&{}\quad{} + \lambda_U^{\dagger} \lambda_U  - \frac{16}{3} g_3^2 - 3 g_2^2 - \frac{7}{15} g_1^2 \Bigr] \nonumber \\
&{}\quad{} + 2 \hat{\lambda}_D \Bigl[ \tr (3 \lambda_D^{\dagger} A_D + 3 {\hat{\lambda}_D}^{\dagger} \hat{A}_D + \lambda_E^{\dagger} A_E + {\hat{\lambda}_E}^{\dagger} \hat{A}_E) + 2 \lambda_D^{\dagger} A_D + 2 {\hat{\lambda}_D}^{\dagger} \hat{A}_D \nonumber \\
&{}\quad{} + \lambda_U^{\dagger} A_U + \frac{16}{3} M_3 g_3^2 + 3 M_2 g_2^2 + \frac{7}{15} M_1 g_1^2 \Bigr]   
\mytag \\%
(4 \pi)^2 \frac{\text{d} \hat{A}_E}{\text{d} t} &= \hat{A}_E \Bigl[ \tr (3 \lambda_D^{\dagger} \lambda_D + 3 {\hat{\lambda}_D}^{\dagger} \hat{\lambda}_D + \lambda_E^{\dagger} \lambda_E +  {\hat{\lambda}_E}^{\dagger} \hat{\lambda}_E) + 5 {\hat{\lambda}_E}^{\dagger} \hat{\lambda}_E - 3 g_2^2 - \frac{9}{5} g_1^2 \Bigr] \nonumber \\
&{}\quad{} + 2 \hat{\lambda}_E \Bigl[ \tr (3 \lambda_D^{\dagger} A_D + 3 {\hat{\lambda}_D}^{\dagger} \hat{A}_D + \lambda_E^{\dagger} A_E + {\hat{\lambda}_E}^{\dagger} \hat{A}_E) + 2 {\hat{\lambda}_E}^{\dagger} \hat{A}_E \nonumber \\
&{}\quad{}  + 3 M_2 g_2^2 + \frac{9}{5} M_1 g_1^2 \Bigr] + 5 A_E {\lambda_E}^{\dagger} \hat{\lambda}_E + 4 \lambda_E {\lambda_E}^{\dagger} \hat{A}_E
\mytag 
\end{align}

\subsection[The $B\mu$ term and other bilinear SUSY breaking parameters]{The $\boldsymbol{B\mu}$ term and other bilinear SUSY breaking parameters}
The running of the dimension 2 coefficients of the holomorphic terms in the soft breaking Lagrangian is given by
\globallabel{5}
\begin{align}
(4 \pi)^2 \frac{\text{d} B}{\text{d} t} &= B \Bigl[ \tr (3 \lambda_U^{\dagger} \lambda_U + 3 \lambda_D^{\dagger} \lambda_D + 3 {\hat{\lambda}_D}^{\dagger} \hat{\lambda}_D + \lambda_E^{\dagger} \lambda_E +  {\hat{\lambda}_E}^{\dagger} \hat{\lambda}_E) - 3 g_2^2 - \frac{3}{5} g_1^2 \Bigr] \nonumber \\ 
&{}\quad{} + 2 \mu \Bigl[ \tr (3 \lambda_U^{\dagger} A_U + 3 \lambda_D^{\dagger} A_D + 3 {\hat{\lambda}_D}^{\dagger} \hat{A}_D + \lambda_E^{\dagger} A_E + {\hat{\lambda}_E}^{\dagger} \hat{A}_E) \nonumber \\
&{}\quad{}  + 3 M_2 g_2^2 + \frac{3}{5} M_1 g_1^2 \Bigr]   
\mytag \\%
(4 \pi)^2 \frac{\text{d} B_{D}}{\text{d} t} &= 2 \hat{\lambda}_D  \Bigl( {\hat{\lambda}_D}^{\dagger} B_{D} + \lambda_D^{\dagger} B_{d D} \Bigr) + 4 \hat{A}_D  \Bigl( {\hat{\lambda}_D}^{\dagger} M_{D} + \lambda_D^{\dagger} M_{d D} \Bigr) \nonumber \\
&{}\quad{}  - B_{D} \Bigl( \frac{16}{3} g_3^2 + \frac{4}{15} g_1^2 \Bigr) + M_{D} \Bigl( \frac{32}{3} M_3 g_3^2 + \frac{8}{15} M_1 g_1^2 \Bigr) 
\mytag \\%
(4 \pi)^2 \frac{\text{d} B_{d D}}{\text{d} t} &= 2 \lambda_D  \Bigl( {\hat{\lambda}_D}^{\dagger} B_{D} + \lambda_D^{\dagger} B_{d D} \Bigr) + 4 A_D  \Bigl( {\hat{\lambda}_D}^{\dagger} M_{D} + \lambda_D^{\dagger} M_{d D} \Bigr) \nonumber \\ 
&{}\quad{}  - B_{d D} \Bigl(\frac{16}{3} g_3^2 + \frac{4}{15} g_1^2 \Bigr) + M_{d D} \Bigl(\frac{32}{3} M_3 g_3^2 + \frac{8}{15} M_1 g_1^2 \Bigr) 
\mytag \\%
(4 \pi)^2 \frac{\text{d} B_L}{\text{d} t} &= \hat{\lambda}_E^T \Bigl( \lambda_E^* B_{lL} + \hat{\lambda}_E^* B_L \Bigr)  + 2 \hat{A}_E^T \Bigl( \lambda_E^* M_{lL} + \hat{\lambda}_E^* M_L \Bigl) \nonumber \\
&{}\quad{}  - B_L \Bigl(3 g_2^2 + \frac{3}{5} g_1^2 \Bigr) + M_L \Bigl(6 M_2 g_2^2 + \frac{6}{5} M_1 g_1^2 \Bigr) 
\mytag \\%
(4 \pi)^2 \frac{\text{d} B_{l L}}{\text{d} t}  &= \lambda_E^T \Bigl( \lambda_E^* B_{lL} + \hat{\lambda}_E^* B_L \Bigr)  + 2 A_E^T \Bigl( \lambda_E^* M_{lL} + \hat{\lambda}_E^* M_L \Bigl) \nonumber \\ 
&{}\quad{}  - B_{l L} \Bigl(3 g_2^2 + \frac{3}{5} g_1^2 \Bigr) + M_{l L} \Bigl(6 M_2 g_2^2 + \frac{6}{5} M_1 g_1^2 \Bigr) \,.
\mytag 
\end{align}

\subsection{Soft scalar masses}

Finally we study the running of the sfermion and Higgs masses parameters.
It is convenient to define the quantity
\begin{equation}
\mathcal{S} = m^2_{h_u} - m^2_{h_d} + \tr ( m^2_q - 2 m^2_{u^c} + m^2_{d^c} - m^2_{l} + m^2_{e^c} + m^2_{D^c} - m^2_{\overline{D^c}} - m^2_{L} + m^2_{\overline{L}} )\,. 
\end{equation}
As usual, below the scale where a degree of freedom is integrated out the corresponding entries in the $m^2$ matrices will vanish in $\mathcal{S}$ and in the equations below. The RGE equations are then
\globallabel{6}
\begin{align}
(4 \pi)^2 \frac{\text{d} m^2_{h_u}}{\text{d} t} &= 6 \tr \Bigl( (m^2_{h_u} + m^2_q) \lambda_U^{\dagger} \lambda_U + \lambda_U^{\dagger} m^2_{u^c} \lambda_U + A_U^{\dagger} A_U \Bigr) \nonumber \\
&{}\quad{}  - 6 |M_2|^2 g_2^2 - \frac{6}{5} |M_1|^2 g_1^2 + \frac{3}{5} g_1^2 \mathcal{S}  
\mytag \\%
(4 \pi)^2 \frac{\text{d} m^2_{h_d}}{\text{d} t} &= \tr \Bigl( 6 (m^2_{h_d} + m^2_q) \lambda_D^{\dagger} \lambda_D + 6 (m^2_{h_d} + m^2_q) {\hat{\lambda}_D}^{\dagger} \hat{\lambda}_D  + 2 (m^2_{h_d} + m^2_{l}) {\lambda_E}^{\dagger} \lambda_E \nonumber \\
&{}\quad{}  + 2 (m^2_{h_d} + m^2_{L}) {\hat{\lambda}_E}^{\dagger} \hat{\lambda}_E + 2 \lambda_E^{\dagger} \hat{\lambda}_E m^2_{l L} + 2 {\hat{\lambda}_E}^{\dagger} \lambda_E {m^2_{l L}}^{\dagger} + 6 \lambda_D^{\dagger} m^2_{d D} \hat{\lambda}_D \nonumber \\
&{}\quad{}  + 6 {\hat{\lambda}_D}^{\dagger} {m^2_{d D}}^{\dagger} \lambda_D + 6 \lambda_D^{\dagger} m^2_{d^c} \lambda_D + 6 {\hat{\lambda}_D}^{\dagger} m^2_{D^c} \hat{\lambda}_D + 2 \lambda_E^{\dagger} m^2_{e^c} \lambda_E + 2 {\hat{\lambda}_E}^{\dagger} m^2_{e^c} \hat{\lambda}_E \Bigr) \nonumber \\
&{}\quad{}  + 2 \tr \Bigl(  3 A_D^{\dagger} A_D + 3 {\hat{A}_D}^{\dagger} \hat{A}_D + A_E^{\dagger} A_E + {\hat{A}_E}^{\dagger} \hat{A}_E \Bigr) \nonumber \\
&{}\quad{} - 6 |M_2|^2 g_2^2 - \frac{6}{5} |M_1|^2 g_1^2 - \frac{3}{5} g_1^2 \mathcal{S}  
\mytag \\%
(4 \pi)^2 \frac{\text{d} m^2_{q}}{\text{d} t} &=  (m^2_q + 2 m^2_{h_u}) \lambda_U^{\dagger} \lambda_U + (m^2_q + 2 m^2_{h_d}) (\lambda_D^{\dagger} \lambda_D + {\hat{\lambda}_D}^{\dagger} \hat{\lambda}_D) \nonumber \\
&{}\quad{}  + (\lambda_U^{\dagger} \lambda_U + \lambda_D^{\dagger} \lambda_D + {\hat{\lambda}_D}^{\dagger} \hat{\lambda}_D) m^2_q + 2 (A_U^{\dagger} A_U + A_D^{\dagger} A_D + {\hat{A}_D}^{\dagger} \hat{A}_D) \nonumber \\  
&{}\quad{} + 2 (\lambda_U^{\dagger} m^2_{u^c} \lambda_U + \lambda_D^{\dagger} m^2_{d^c} \lambda_D + {\hat{\lambda}_D}^{\dagger} m^2_{D^c} \hat{\lambda}_D + {\hat{\lambda}_D}^{\dagger} {m^2_{d D}}^{\dagger} \lambda_D +  {\lambda_D}^{\dagger} m^2_{d D} \hat{\lambda}_D) \nonumber \\
&{}\quad{} - \frac{32}{3} |M_3|^2 g_3^2 - 6 |M_2|^2 g_2^2 - \frac{2}{15} |M_1|^2 g_1^2 + \frac{1}{5} g_1^2 \mathcal{S}  
\mytag \\%
(4 \pi)^2 \frac{\text{d} m^2_{l}}{\text{d} t} &= (m^2_{l} + 2 m^2_{h_d}) \lambda_E^{\dagger} \lambda_E + {m^2_{l L}}^{\dagger} {\hat{\lambda}_E}^{\dagger} \lambda_E + \lambda_E^{\dagger} \lambda_E m^2_{l} + \lambda_E^{\dagger} \hat{\lambda}_E m^2_{l L} \nonumber \\
&{}\quad{} + 2 \lambda_E^{\dagger} m^2_{e^c} \lambda_E + 2 A_E^{\dagger} A_E - 6 |M_2|^2 g_2^2 - \frac{6}{5} |M_1|^2 g_1^2 - \frac{3}{5} g_1^2 \mathcal{S} 
\mytag \\%
(4 \pi)^2 \frac{\text{d} m^2_{u^c}}{\text{d} t} &= 2 (m^2_{u^c} + 2 m^2_{h_u}) \lambda_U \lambda_U^{\dagger} + 2 \lambda_U \lambda_U^{\dagger} m^2_{u^c} + 4 \lambda_U m^2_q \lambda_U^{\dagger} + 4 A_U A_U^{\dagger} \nonumber \\
&{}\quad{}  - \frac{32}{3} |M_3|^2 g_3^2 - \frac{32}{15} |M_1|^2 g_1^2 - \frac{4}{5} g_1^2 \mathcal{S}  
\mytag \\%
(4 \pi)^2 \frac{\text{d} m^2_{d^c}}{\text{d} t} &= 2 (m^2_{d^c} + 2 m^2_{h_d}) \lambda_D \lambda_D^{\dagger} + 2 m^2_{d D} \hat{\lambda}_D \lambda_D^{\dagger} + 2 \lambda_D \lambda_D^{\dagger} m^2_{d^c} + 2 \lambda_D {\hat{\lambda}_D}^{\dagger} {m^2_{d D}}^{\dagger} \nonumber \\
&{}\quad{} + 4 \lambda_D m^2_q \lambda_D^{\dagger} + 4 A_D A_D^{\dagger} - \frac{32}{3} |M_3|^2 g_3^2 - \frac{8}{15} |M_1|^2 g_1^2 + \frac{2}{5} g_1^2 \mathcal{S}  
\mytag \\%
(4 \pi)^2 \frac{\text{d} m^2_{e^c}}{\text{d} t} &= 2 (m^2_{e^c} + 2 m^2_{h_d}) (\lambda_E \lambda_E^{\dagger} + \hat{\lambda}_E {\hat{\lambda}_E}^{\dagger}) + 2 (\lambda_E \lambda_E^{\dagger} + \hat{\lambda}_E {\hat{\lambda}_E}^{\dagger}) m^2_{e^c} \nonumber \\
&{}\quad{} + 4 (\lambda_E m^2_{l} \lambda_E^{\dagger} + \hat{\lambda}_E m^2_{L} {\hat{\lambda}_E}^{\dagger} + \lambda_E {m^2_{l L}}^{\dagger} {\hat{\lambda}_E}^{\dagger} + \hat{\lambda}_E m^2_{l L} \lambda_E^{\dagger}) + 4 (A_E A_E^{\dagger} + \hat{A}_E {\hat{A}_E}^{\dagger}) \nonumber \\
&{}\quad{} - \frac{24}{5} |M_1|^2 g_1^2 + \frac{6}{5} g_1^2 \mathcal{S}  
\mytag \\%
(4 \pi)^2 \frac{\text{d} m^2_{D^c}}{\text{d} t} &= 2 (m^2_{D^c} + 2 m^2_{h_d}) \hat{\lambda}_D {\hat{\lambda}_D}^{\dagger} + 2 {m^2_{d D}}^{\dagger} \lambda_D {\hat{\lambda}_D}^{\dagger} + 2 \hat{\lambda}_D {\hat{\lambda}_D}^{\dagger} m^2_{D^c} + 2 \hat{\lambda}_D \lambda_D^{\dagger} m^2_{d D} \nonumber \\
&{}\quad{} + 4 \hat{\lambda}_D m^2_q {\hat{\lambda}_D}^{\dagger} + 4 \hat{A}_D {\hat{A}_D}^{\dagger} - \frac{32}{3} |M_3|^2 g_3^2 - \frac{8}{15} |M_1|^2 g_1^2 + \frac{2}{5} g_1^2 \mathcal{S}  
\mytag \\%
(4 \pi)^2 \frac{\text{d} m^2_{\overline{D^c}}}{\text{d} t} &= - \frac{32}{3} |M_3|^2 g_3^2 - \frac{8}{15} |M_1|^2 g_1^2 - \frac{2}{5} g_1^2 \mathcal{S} 
\mytag \\%
(4 \pi)^2 \frac{\text{d} m^2_{d D}}{\text{d} t} &= 2 (m^2_{d^c} + 2 m^2_{h_d}) \lambda_D \hat{\lambda}_D^{\dagger} + 2 \lambda_D \lambda_D^{\dagger} m^2_{d D} + 2 \lambda_D \hat{\lambda}_D^{\dagger} m^2_{D^c} + 2 m^2_{d D} \hat{\lambda}_D {\hat{\lambda}_D}^{\dagger} \nonumber \\
&{}\quad{} + 4 \lambda_D m^2_q \hat{\lambda_D}^{\dagger} + 4 A_D \hat{A}_D^{\dagger}  
\mytag \\%
(4 \pi)^2 \frac{\text{d} m^2_{\overline{L}}}{\text{d} t} &= - 6 |M_2|^2 g_2^2 - \frac{6}{5} |M_1|^2 g_1^2 + \frac{3}{5} g_1^2 \mathcal{S}  
\mytag \\%
(4 \pi)^2 \frac{\text{d} m^2_{L}}{\text{d} t} &= (m^2_{L} + 2 m^2_{h_d}) {\hat{\lambda}_E}^{\dagger} \hat{\lambda}_E + m^2_{l L} \lambda_E^{\dagger} \hat{\lambda}_E + {\hat{\lambda}_E}^{\dagger} \hat{\lambda}_E m^2_{L} + {\hat{\lambda}_E}^{\dagger} \lambda_E {m^2_{l L}}^{\dagger} \nonumber \\
&{}\quad{} + 2 {\hat{\lambda}_E}^{\dagger} m^2_{e^c} \hat{\lambda}_E + 2 {\hat{A}_E}^{\dagger} \hat{A}_E  - 6 |M_2|^2 g_2^2 - \frac{6}{5} |M_1|^2 g_1^2 - \frac{3}{5} g_1^2 \mathcal{S} 
\mytag \\%
(4 \pi)^2 \frac{\text{d} m^2_{l L}}{\text{d} t} &= (m^2_{L} + 2 m^2_{h_d}) \hat{\lambda}_E^{\dagger} \lambda_E + m^2_{l L} \lambda_E^{\dagger} \lambda_E + \hat{\lambda}_E^{\dagger} \lambda_E m^2_{l} + \hat{\lambda}_E^{\dagger} \hat{\lambda}_E m^2_{l L} \nonumber \\
&{}\quad{} + 2 \hat{\lambda}_E^{\dagger} m^2_{e^c} \lambda_E + 2 \hat{A}_E^{\dagger} A_E \,.
\mytag 
\end{align}

\subsection{Approximate analytical running of Higgs mass parameters}
\label{sec:higgsrun}

A sometimes useful simple approximation for the solutions of the RGEs for the soft mass terms is obtained in the limit in which $\tan\beta$ is moderate, so that only the top Yukawa coupling is relevant in the equations above, and the squared gaugino masses and $A$-terms are negligible compared to $m^2_{10}$. In such a case, the only soft terms that run significantly are $m^2_{h_u}$ and the stop squared mass parameters ${m}^2_{q_3}$ and ${m}^2_{u^c_3}$, for which we have
(see, e.g.\ appendix of \cite{Barbieri:1995tw})
\begin{equation}
\label{eq:running}
\begin{aligned}
m^2_{h_u}(M^2_Z) &= m^2_{h_u}(M_\text{GUT}) - \frac{1}{2} m^2_U \, \rho 
= -\frac{1}{2} m_{10}^2 \, (4+5(-2+\rho)\sin^2\theta_u) \\
{m}^2_{q_3}(M^2_Z) &= {m}^2_{q_3}(M_\text{GUT}) - \frac{1}{6} m^2_U \, \rho 
= m^2_{10} \Big( 1 -\frac{5}{6} \rho\, \sin^2\theta_u \Big) \\
{m}^2_{u^c_3}(M^2_Z) &= {m}^2_{u^c_3}(M_\text{GUT}) - \frac{1}{3} m^2_U \, \rho
= m^2_{10} \Big( 1 -\frac{5}{3} \rho\, \sin^2\theta_u \Big)\,,
\end{aligned}
\end{equation}
where $m^2_U = (m^2_{h_u}+{m}^2_{q_3}+{m}^2_{u^c_3})_{M_\text{GUT}} = 5 \sin^2\theta_u \,m^2_{10}$, $m^2_{h_u}(M_\text{GUT}) = (-2 \cos^2\theta_u + 3 \sin^2\theta_u)m^2_{10}$, ${m}^2_{q_3}(M_\text{GUT}) = {m}^2_{u^c_3}(M_\text{GUT}) = m^2_{10}$ and 
\begin{equation}
\label{eq:rho}
\rho = 1-\exp\left({\displaystyle 12\int \frac{dt}{(4\pi)^2} \lambda^2_t(t)}\right)\,, \quad 0 < \rho < 1 \,.
\end{equation}
A typical value of $\rho$ is $\rho \sim 0.7$.

\section{Razor}
\label{app:razor}

The razor analysis \cite{razorpaper} is a fairly recent approach that has been introduced by the CMS collaboration to discriminate New Physics signals over SM backgrounds in situations in which there is a presence of large $E_T^{\text{miss}}$. The framework is designed to perfectly fit to a situation in which from parton collisions two heavy particles ($G_1$, $G_2$), whose mass is significantly larger than those of SM particles, are produced. The decays of the $G_i$'s are then forced to be described by a dijet topology, in which any of the $G_i$ decays to a massive unseen particle $\chi_i$, contributing to $E_T^{\text{miss}}$, and a massless seen particle $Q_i$, being detected as a jet. In SUSY theories the benchmark scenario for this approach would thus be the case in which two heavy squarks are produced and then decay to a quark and a neutralino:
\begin{equation}
pp \rightarrow G_1 G_2 \rightarrow Q_1  \chi_1 + Q_2 \chi_2 \Longrightarrow pp \rightarrow \widetilde{q} \widetilde{q} \rightarrow 2 j + \text{MET} \,.
\end{equation}
For any of the decay chains $G_i \rightarrow Q_i + \chi_i$ one can define the variable
\begin{equation}
\label{eq:rzr_Mdelta}
M_{\Delta_i} = \frac{M_{G_i}^2-M_{\chi_i}^2}{M_{G_i}} \,,
\end{equation}
which, in the approximation where the heavy $G_i$'s are produced at threshold and the $Q_i$'s are massless, corresponds to twice the energy of the $Q_i$'s in the center of mass (CM) frame. 

The reconstruction of the CM frame in events with two undetected particles is not conceivable, but still it is possible to perform an event by event reconstruction of the specific reference frame in which the three-momenta of the observed jets coincide. This reference frame, named R-frame, is an estimator of the CM frame itself: working in it one can construct a transverse mass $M^R_T$,
\begin{equation}
M^R_T \equiv \sqrt{\frac{E_T^{miss}(p_T^{j_1}+p_T^{j_1})-\overrightarrow{E}_T^{miss}(\overrightarrow{p}_T^{j_1}+\overrightarrow{p}_T^{j_1})}{2}} \, ,
\end{equation}
whose distribution would have an edge at $M_\Delta$ corresponding to the case in which CM and R frame coincide, and 
\begin{equation}
M_R \equiv \sqrt{(E_{j_1}+E_{j_2})^2-(p_z^{j_1}+p_z^{j_1})^2} \, ,
\end{equation}
which peaks at $M_\Delta$ for signal events. 

Given the tools described one could easily discriminate between background and signal events by means of the razor variable, defined as
\begin{equation}
R \equiv \frac{M^R_T}{M_R} \, .
\end{equation}
For signal events the distribution of $R$ peaks around $1/2$, while for any SM background it is quite lower: this allows to discriminate between the two by means of smart cuts on the value of $R$.

\end{document}